\documentclass[aps,pre,reprint,superscriptaddress,showkeys,showpacs]{revtex4-2}
\usepackage{lineno}
\usepackage{amsfonts}
\usepackage{amssymb}
\usepackage{amsmath}
\usepackage{graphicx}
\usepackage{makeidx}
\usepackage{subfig}
\usepackage{tabularx}
\usepackage{dcolumn}

\usepackage{amsthm}

\usepackage[figuresright]{rotating}

\usepackage{subfig}

\usepackage{graphicx}
\usepackage{dcolumn}
\usepackage{bm}
\usepackage{epstopdf}
\usepackage{epsfig}
\usepackage[colorlinks = true,
linkcolor = blue,
urlcolor  = blue,
citecolor = blue,
anchorcolor = blue]{hyperref}

\usepackage{xcolor}

\DeclareMathOperator{\sech}{sech}

\begin{document}
	
	\title{Pearl supratransmission in a boundary-driven two-dimensional 
		nonlinear Schr\"odinger equation with a hole}
	
	\author{Rudy Kusdiantara}
	\email{rudy\_kusdiantara@itb.ac.id}
	
	\affiliation{Industrial and Financial Mathematics Research Group, Institut Teknologi Bandung, Jl.\ Ganesha No.\ 10, Bandung, 40132, Indonesia}
	
	\author{Hadi Susanto}
	\email{hadi.susanto@yandex.com}
	
	\affiliation{Department of Mathematics, Khalifa University, PO Box
		127788,\\ Abu Dhabi, United Arab Emirates}
	
	


	\date{\today}
	
	\begin{abstract}
		We investigate energy supratransmission in a boundary-driven two-dimensional nonlinear Schr\"odinger equation with a central hole. Harmonic forcing with azimuthal modulation generates standing-wave states whose existence and stability depend on the driving amplitude, the inner radius, and the imposed azimuthal charge. Bifurcation analysis shows that small inner radii produce strongly confined states with higher destabilization thresholds, whereas larger radii yield broader profiles	and smoother transitions between stable and unstable branches. The cubic--quintic and saturable models exhibit similar qualitative	behaviour but differ quantitatively in their critical amplitudes and parameter dependence. A variational approximation captures the dependence of the critical drive on the azimuthal charge and nonlinear parameters, and clarifies how the nonlinear response shapes the stationary states near the turning point. Time-dependent simulations show that supratransmission occurs through the emission of localized pulses, with nonzero azimuthal charge triggering symmetry breaking and producing two-dimensional localized excitations (pearls). Isosurface plots provide a complementary view of the resulting radial and angular excursions. These results establish a quantitative framework for supratransmission in	two-dimensional geometries and are relevant to driven nonlinear systems in optics, Bose--Einstein condensates, and structured media.
	\end{abstract}

	\maketitle
	
	
	\section{Introduction}
	
	Nonlinear supratransmission refers to the transmission of energy through a
	frequency band in which linear waves are evanescent. In a linear medium,
	driving at a frequency inside the band gap produces only exponentially
	decaying responses \cite{markos2008wave}, whereas frequencies in the allowed band yield propagating waves. In contrast, nonlinear media
	can sustain energy transfer once the driving amplitude exceeds a critical
	threshold. Above this threshold, nonlinear instabilities or the suppression of the
	evanescent linear response give rise to localized structures that carry
	energy through the forbidden band \cite{caputo2001nonlinear}. The effect was then
	established in the harmonically driven sine--Gordon chain
	\cite{geniet2002energy,geniet2003nonlinear}, where a sharp transition in
	transmission occurs at a well-defined critical amplitude.
	
	Supratransmission has since been observed in a broad range of discrete and
	continuous systems. Mechanical lattices of coupled pendula display the
	formation of breather-like excitations at the onset of energy
	transmission~\cite{geniet2003nonlinear}, while optical media with Kerr
	nonlinearity support analogous mechanisms for energy flow through forbidden
	frequency ranges~\cite{khomeriki2004nonlinearband}. Experimental realizations in electrical transmission lines~\cite{koon2014experimental}, granular chains~\cite{lydon2015nonlinear}, and origami-based metamaterials \cite{zhang2020programmable,wang2023highly} confirm the theoretical
	predictions and highlight potential applications in controlled energy
	transport. Systems with nearly flat bands also exhibit sharp
	amplitude-dependent transitions in transmission properties
	\cite{susanto2023surge,kusdiantara2025band}.
	
	Higher-dimensional settings introduce additional geometric and boundary
	constraints that influence the onset and the nature of supratransmission.
	Two-dimensional crystal lattices can support discrete breathers
	\cite{medvedev2011localized}, while three-dimensional nonlinear
	Klein--Gordon systems display threshold-driven energy penetration mediated
	by localized excitations~\cite{macias2008numerical2,macias2005numerical}.
	Studies on intermetallic alloys, including Pt$_3$Al, show that spatial
	dimensionality and domain geometry play essential roles in the stability
	and propagation of localized modes
	\cite{cherednichenko2019nonlinear,zakharov2023effect}. Abdullina et al.\ \cite{abdullina2026linear} demonstrate supratransmission in a three-dimensional B2 (CsCl-type) lattice and show a sublattice-selective mechanism in which gap and above-spectrum breathers propagate preferentially along heavy and light atomic sublattices, respectively, highlighting the role of mass contrast and lattice structure in nonlinear energy transport. 
	
	In this work, we examine supratransmission in a two-dimensional nonlinear
	Schr\"odinger equation (NLS) defined on a disk with a central hole of
	radius~$r_{\min}$. The system is driven harmonically on the inner ring with an
	azimuthal modulation of topological charge~$m$. This configuration enables
	the study of supratransmission in the presence of angular momentum, an
	aspect that has received little attention. We show that a critical driving
	amplitude exists above which energy propagates through the medium via
	two-dimensional solitary structures. Numerical simulations and a
	variational approximation quantify how the critical amplitude depends on
	the driving frequency, the hole radius, and the azimuthal charge. A
	stability analysis reveals how geometry, nonlinearity, and rotational
	structure interact to control the onset of transmission.
	
	The results extend supratransmission theory to two-dimensional NLS models
	with nontrivial geometry and may inform applications in nonlinear optics,
	acoustic and elastic metamaterials, and controlled energy transport.
	In particular, the proposed setup enables supratransmission to act as a
	robust mechanism for creating localized Townes solitons (that we refer to as pearls here) as well as ring-shaped ones.
	
	The paper is organized as follows. Section~\ref{Sec:Model} introduces the mathematical formulation and the definition of supratransmission. Section~\ref{Sec:Stand} analyzes localized standing waves and their stability, and presents a variational approximation for the critical threshold. Section~\ref{sec:dynamics} illustrates the dynamics beyond the critical driving amplitude. Section~\ref{sec:allowed} examines the nonlinear response when the driving frequency lies within the allowed band. In this regime, we
	identify an energy jump occurring above a critical amplitude, analogous to supratransmission but arising within the allowed band. Section~\ref{Sec:Conclusions} concludes the main findings.

	\section{Mathematical Model and Supratransmission}
	\label{Sec:Model}
	
	\begin{figure}[tbhp!] 
		\centering 
		\includegraphics[scale=.55]{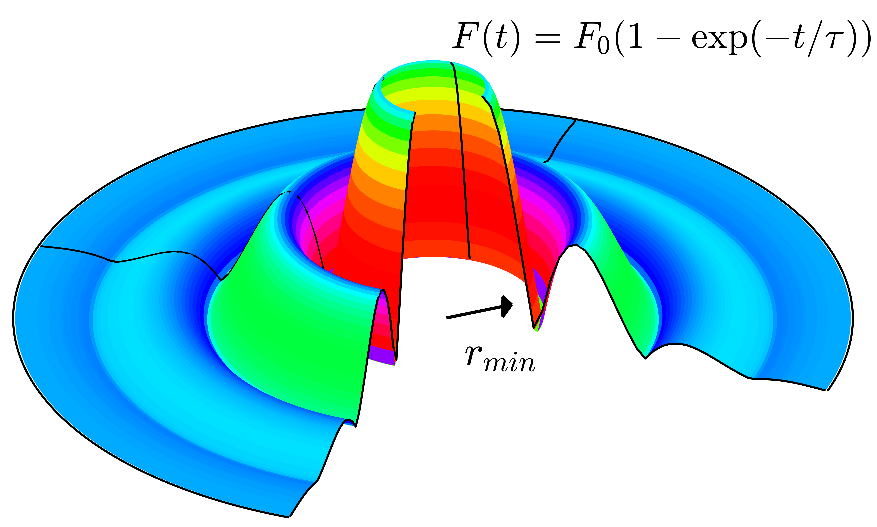} 
		\caption{
			Schematic of the annular geometry used for the driven (2+1)-dimensional
			nonlinear Schr\"odinger equation. Harmonic forcing with azimuthal phase
			$e^{im\theta}$ is applied at the inner boundary $r=r_{\min}$, injecting
			angular momentum when $m\neq 0$. The function $F(t)$ denotes the
			time-dependent driving amplitude. 
		}
		\label{fig:ilus} 
	\end{figure}

	We consider a boundary-driven (2+1)-dimensional nonlinear Schr\"odinger
	equation written in polar coordinates,
	\begin{equation}
		iu_t = \left(\partial_{rr} + \frac{1}{r}\partial_r +
		\frac{1}{r^2}\partial_{\theta\theta}\right)u + N(|u|^2)\,u,
		\label{main1}
	\end{equation}
	posed on the annular region $r \in [r_{\min},\infty)$, $\theta\in[0,2\pi)$.
	The function $N(|u|^2)$ represents the nonlinear response of the medium
	and determines the structure and stability of stationary and driven
	states.
	
	A purely cubic nonlinearity $N(|u|^2)=2|u|^2$ is known to be
	scale-invariant and may lead to finite-time collapse. To avoid this
	behaviour, we employ two modified nonlinearities that regularize the
	high-intensity response. The first is the cubic--quintic model,
	\begin{equation}
		N(|u|^2)=2|u|^2 - s|u|^4,\label{nlin:cq}
	\end{equation}
	where $s>0$ introduces a defocusing quintic term. The second is a
	saturable model,
	\begin{equation}
		N(|u|^2)=\frac{2|u|^2}{1+\frac{s}2|u|^2}, \label{nlin:sat}
	\end{equation}
	which saturates at large amplitudes. The two models exhibit different stability characteristics. In the
	cubic--quintic case, the defocusing quintic term can stabilize both
	radially symmetric solitary waves and certain vortex states \cite{pego2002spectrally,michinel2001excitation,mihalache2002stable,malomed2002stability}. In
	saturable media, collapse is suppressed by saturation, but vortex-type
	profiles may remain susceptible to azimuthal modulation instabilities
	depending on the saturation level and mode number \cite{skryabin1998dynamics}. In what follows, the cubic--quintic model is used as the primary example, with remarks on the saturable case included where the behaviour differs.
	
	The system is driven at the inner boundary $r=r_{\min}$ by prescribing
	\begin{equation}
		u(r_{\min},t) = F(t)\, e^{i m\theta - i\Omega t},
		\label{drive}
	\end{equation}
	where $\Omega$ is the driving frequency and
	$m\in\mathbb{Z}$ is the azimuthal index of the imposed phase. 
	A schematic of the annular geometry and boundary forcing is shown in Fig.~\ref{fig:ilus}.
	The amplitude evolves according to
	\begin{equation}
		F(t) = F_0\big(1-e^{-t/\tau}\big), \qquad \tau\gg1,
		\label{eq:inc}
	\end{equation}
	so that the forcing increases adiabatically. This boundary condition
	injects angular momentum when $m\neq 0$ and provides a controlled means
	of exciting standing and traveling structures inside the annulus.

	\section{Analysis of radially localized solutions}
	\label{Sec:Stand}
	
	Radially symmetric standing waves of Eq.~\eqref{main1} are obtained by
	substituting the ansatz
	\[
	u(r,\theta,t) = U(r)e^{im\theta - i\Omega t},
	\]
	where $m\in\mathbb{Z}$ is the topological charge and $\Omega$ is the
	nonlinear frequency. This reduces Eq.~\eqref{main1} to the stationary
	radial equation
	\begin{equation}
		\left(\partial_{rr}+\frac{1}{r}\partial_r-\frac{m^2}{r^2}\right)U
		-\Omega U + N(U^2)U = 0,
		\label{main2}
	\end{equation}
	supplemented with the inner boundary condition $U(r_{\min})=F_0$. 
	
	\subsection{Spectral band}
	
	First, we determine the linear band structure. Setting $N\equiv 0$ in
	Eq.~\eqref{main2} yields the linear equation
	\[
	U'' + \frac{1}{r}U' - \left(\frac{m^{2}}{r^{2}}+\Omega\right)U = 0,
	\]
	which is a Bessel-type equation with spectral parameter $\Omega$. Writing
	$k^{2}=-\Omega$, we see that for $\Omega<0$ one has $k\in\mathbb{R}$ and
	the solutions are oscillatory Bessel functions $J_m(kr)$ and $Y_m(kr)$,
	corresponding to propagating cylindrical waves \cite{abramowitz1964handbook}. 
	Thus the allowed band of the linear problem occupies $\Omega\le 0$.
	
	For $\Omega>0$, we have $k=i\sigma$ with $\sigma=\sqrt{\Omega}>0$, and
	the solutions are the modified Bessel functions $I_m(\sigma r)$ and
	$K_m(\sigma r)$, which grow or decay exponentially with radius. In this
	regime no propagating linear modes exist and the field is purely
	evanescent. Hence, for the sign convention in Eq.~\eqref{main1}, the
	interval $\Omega>0$ plays the role of a forbidden band (band gap) for
	linear wave propagation, and supratransmission corresponds to nonlinear
	energy transport through this linearly prohibited frequency range.
	
	{In the simulations below, we consider the driving frequencies $\Omega=1$. According to the spectral classification above, the value of $\Omega$ lies in the evanescent (forbidden) regime where the linear solutions decay
		exponentially.} 
	
	\subsection{Nonlinear solutions}
	
	We are interested in radially localized nonlinear solutions of \eqref{main2} in the band gap region that appear due to the drive. Once a solution $U(r)$ is obtained, we investigate its stability by perturbing it with angular modes of wavenumber~$q$ of the form
	\begin{equation}
		u = \Big(U(r) +
		a(r)e^{iq\theta+\lambda t} + b^*(r)e^{-iq\theta+\lambda^* t}\Big)
		e^{im\theta - i\Omega t},
		\label{stabansatz}
	\end{equation}
	where $a$ and $b$ are small perturbations and $\lambda$ is the
	growth rate. Substituting \eqref{stabansatz} into Eq.~\eqref{main1} and linearizing yields the coupled eigenvalue problem 
	\begin{equation}
		\begin{aligned}
			&\left(\partial_{rr}+\frac{1}{r}\partial_r
			-\frac{(q+m)^2}{r^2}\right)a - \Omega a
			+ N(U^2)a \\
			&\qquad\qquad + U^2 N_\xi(U^2)(a+b) = i\lambda a, \\[0.5em]
			&\left(\partial_{rr}+\frac{1}{r}\partial_r
			-\frac{(q-m)^2}{r^2}\right)b - \Omega b
			+ N(U^2)b \\
			&\qquad\qquad + U^2 N_\xi(U^2)(a+b) = -i\lambda b,
		\end{aligned}
		\label{evp}
	\end{equation}
	where $N_\xi(U^2)=\left.\mathrm{d}N(\xi)/\mathrm{d}\xi\right|_{\xi=U^2}$. The perturbations satisfy
	$a(r_{\min})=b(r_{\min})=0$ so that the imposed driving at $r_{\min}$ remains unchanged.
	A standing wave $U(r)$ is spectrally stable if
	$\mathrm{Re}(\lambda)=0$ for every integer $q$. This criterion follows
	from the Hamiltonian structure of the NLS flow and is standard in the
	stability theory of standing waves \cite{kapitula2013spectral}.
	
	Equation~\eqref{main2} is solved numerically by discretizing the radial
	derivatives using central finite differences \cite{leveque1992numerical}
	and applying the Newton--Raphson method to the resulting nonlinear system.
	A Dirichlet condition $U(r_{\max})=0$ is imposed at the truncated outer
	boundary. To follow solution branches across turning points, we employ
	pseudoarclength continuation \cite{keller1987continuation}, which enables
	a systematic exploration of the dependence of standing waves on the
	parameters $(F_0, m)$ and on the nonlinear response $N$.
	
	The eigenvalue problem \eqref{evp} is discretized similarly, producing a
	finite-dimensional matrix system whose eigenpairs $(\lambda,a,b)$ are
	computed using standard eigensolvers. The spectrum provides information
	on the growth or decay of perturbations and identifies the onset of
	azimuthal instabilities. This analysis will help us understand the mechanism behind the dynamical transitions observed under boundary forcing. In particular, the disappearance of localized solutions \cite{susanto2008boundary,hasmi2025supratransmission} or changes in the stability spectrum \cite{susanto2023surge,kusdiantara2025band} can cause supratransmission in our system.

	\subsection{Bifurcation and stability results}
	
	\begin{figure*}[thbp!]
		\centering
		\subfloat[$r_{\min}=0.5$]{\includegraphics[scale=.452]{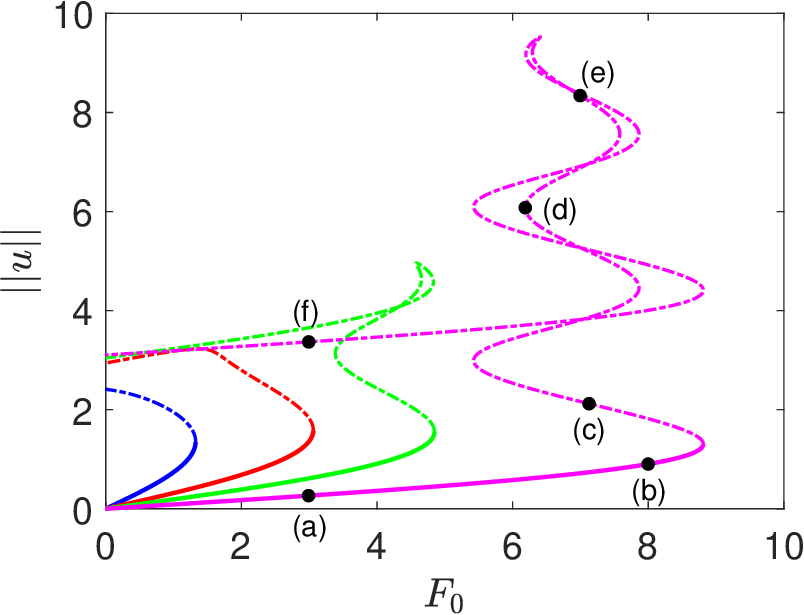}
			\label{subfig:bifur_F0_vs_norm_rm_0_5_s_0}}
		\hspace{0.5em}
		\subfloat[$r_{\min}=2$]{\includegraphics[scale=.452]{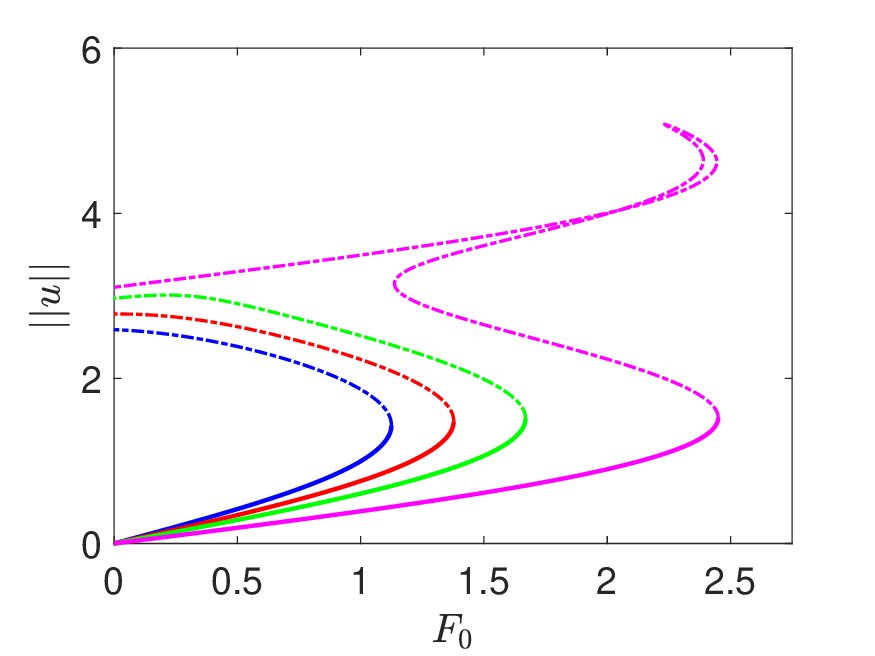}
			\label{subfig:bifur_F0_vs_norm_rm_2_s_0}}
		\caption{
			Bifurcation diagrams for standing waves of Eq.~\eqref{main2} in the
			cubic case ($s=0$), showing the norm as a function of the driving
			amplitude $F_0$ for azimuthal indices $m=0,2,3,5$ (blue, red, green,
			and magenta, respectively). Solid (thick) segments indicate stable
			branches and dashed--dotted (thin) segments unstable ones. Panels (a)
			and (b) correspond to $r_{\min}=0.5$ and $r_{\min}=2$, respectively.
		}
		\label{fig:bifur_F0_vs_norm_rm_0_5_s_0}
	\end{figure*}
	
	Figure~\ref{fig:bifur_F0_vs_norm_rm_0_5_s_0} shows the dependence of the
	norm of standing-wave solutions of Eq.~\eqref{main2} on the driving
	amplitude $F_0$ for the purely cubic model ($s=0$) and azimuthal indices
	$m=0,2,3,5$. The two panels compare inner radii $r_{\min}=0.5$ and $r_{\min}=2$. As
	$F_0$ increases, each branch undergoes saddle-node bifurcations that
	generate alternating stable and unstable segments, with the stability
	information taken from the spectral problem~\eqref{evp}. The structure
	and location of these turning points depend strongly on $m$ and on the
	inner radius. For $r_{\min}=0.5$, the strong geometric confinement produces several
	coexisting stationary branches, corresponding to strongly localized
	states. The lowest branches remain stable over a comparatively broad range of $F_0$. In
	contrast, for $r_{\min}=2$ the influence of the inner boundary is weaker and
	the profiles extend further radially. As a result, the stability
	threshold decreases, and the overall bifurcation structure becomes
	closer to the one-dimensional supratransmission picture, which is
	consistent with the asymptotic reduction of the radial equation as
	$r\to\infty$ \cite{susanto2008boundary}.

	\begin{figure*}[thbp!]
		\centering
		\subfloat[]{\includegraphics[scale=.4]{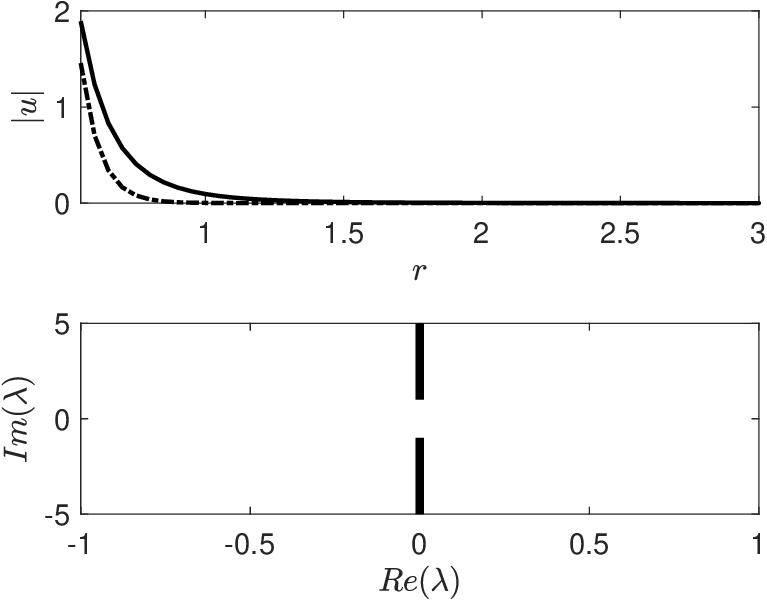}\label{subfig:prof_a}}\hspace{0.5em}
		\subfloat[]{\includegraphics[scale=.4]{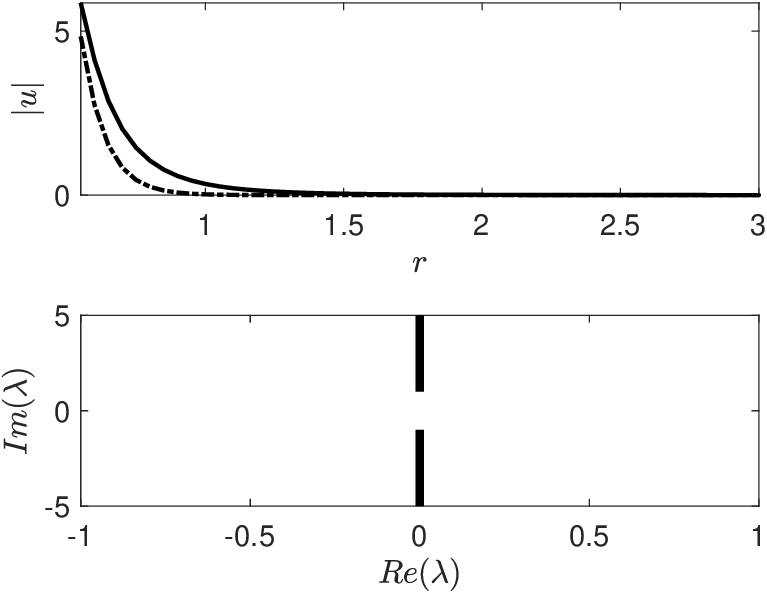}\label{subfig:prof_b}}\hspace{0.5em}
		\subfloat[]{\includegraphics[scale=.4]{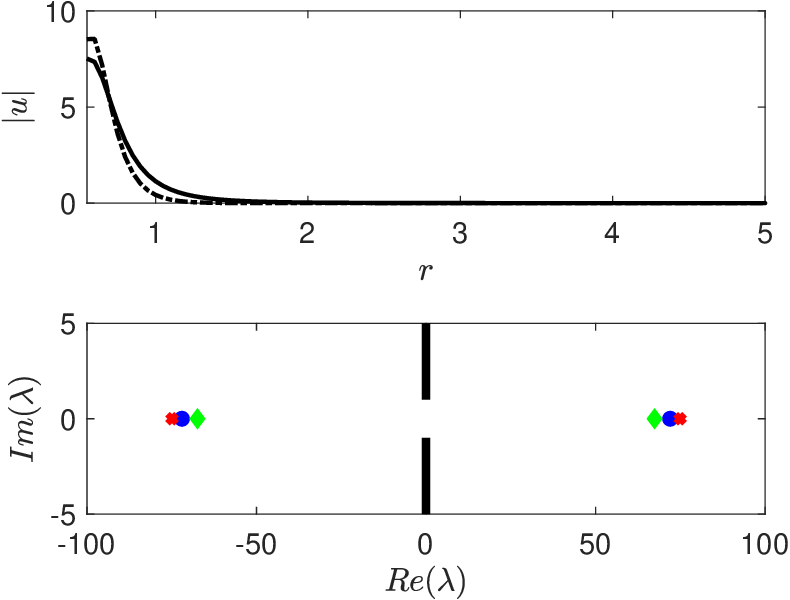}\label{subfig:prof_c}}\\
		\subfloat[]{\includegraphics[scale=.4]{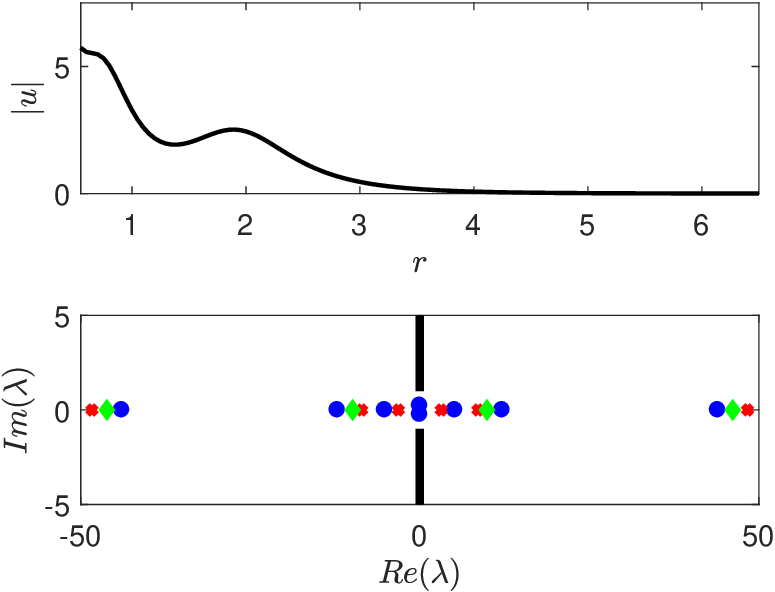}\label{subfig:prof_d}}\hspace{0.5em}
		\subfloat[]{\includegraphics[scale=.4]{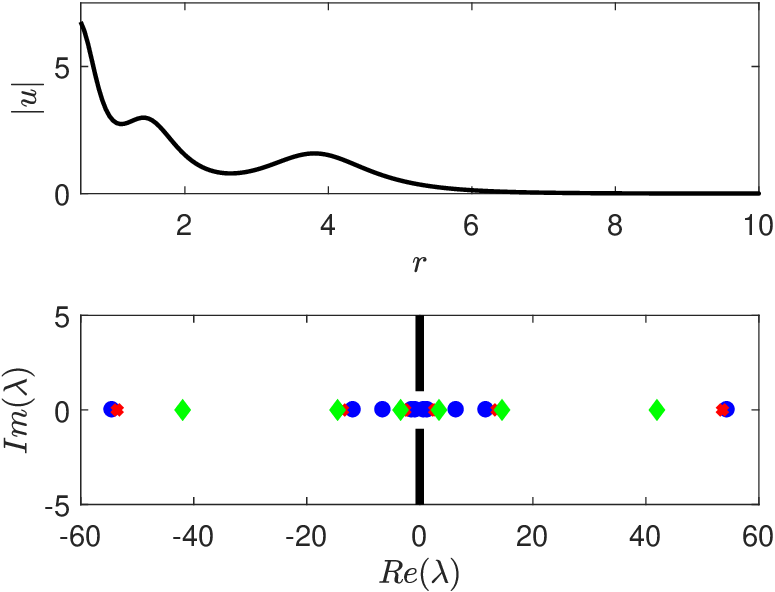}\label{subfig:prof_e}}\hspace{0.5em}
		\subfloat[]{\includegraphics[scale=.4]{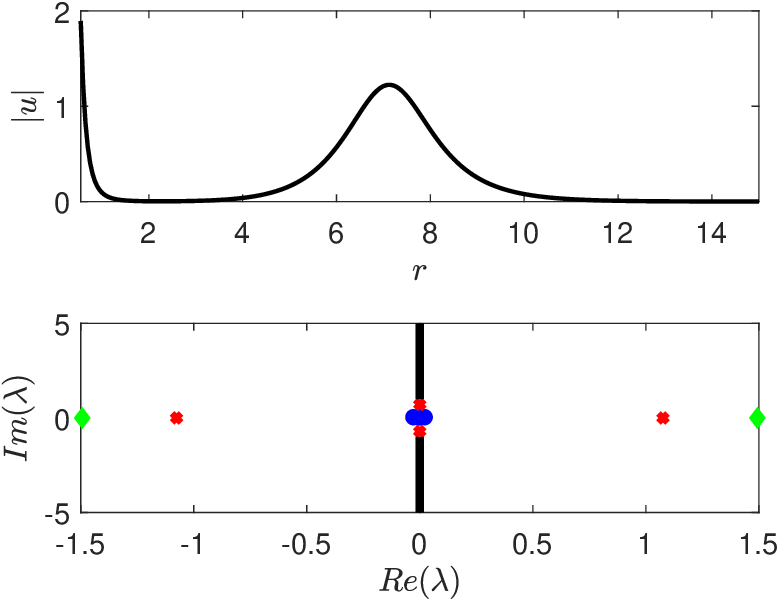}\label{subfig:prof_f}}
		\caption{{
                Selected standing-wave profiles from the bifurcation diagram in Fig.~\ref{subfig:bifur_F0_vs_norm_rm_0_5_s_0} and their corresponding point spectra in the complex plane. The solid curves in the profile plots represent numerical solutions. The dash-dotted curves in panels (a)--(c) show the variational approximation obtained from the ansatz introduced in Eq.~\eqref{var_sep}. Symbols in the spectral plots denote azimuthal perturbation modes: $q=0$ (blue dots), $q=1$ (red crosses), and $q=2$ (green diamonds). Profiles whose spectra lie entirely on the imaginary axis are spectrally stable, whereas eigenvalues with $\mathrm{Re},\lambda \neq 0$ indicate instability.
			}
		}
		\label{fig:prof_bifur}
	\end{figure*}
	
	Figure~\ref{fig:prof_bifur} presents representative profiles sampled from different positions along the bifurcation diagram in Fig.~\ref{subfig:bifur_F0_vs_norm_rm_0_5_s_0}, together with the corresponding point spectra for azimuthal perturbation indices $q=0,1,2$, which correspond to angular Fourier modes. The panels illustrate how changes in $F_0$ affect both spatial localization and spectral properties. Stable profiles are associated with spectra confined to the imaginary axis, whereas instability arises when eigenvalues with nonzero real part emerge, typically due to azimuthal perturbations with $q\ge1$. These comparisons highlight how multidimensional perturbations govern the destabilization of standing waves.
	
	{For clarity of presentation, we display only the first few azimuthal modes ($q=0,1,2$). If these lowest modes already exhibit instability, the solution is spectrally unstable, and it is not necessary to show higher-$q$ modes. Conversely, when the spectra for $q=0,1,2$ lie entirely on the imaginary axis, the computed spectra for higher modes also remain stable in the parameter regimes considered.}

	\begin{figure*}[thbp!]
		\centering
		\subfloat[Cubic--quintic, $r_{\min}=0.5$]{\includegraphics[scale=.452]{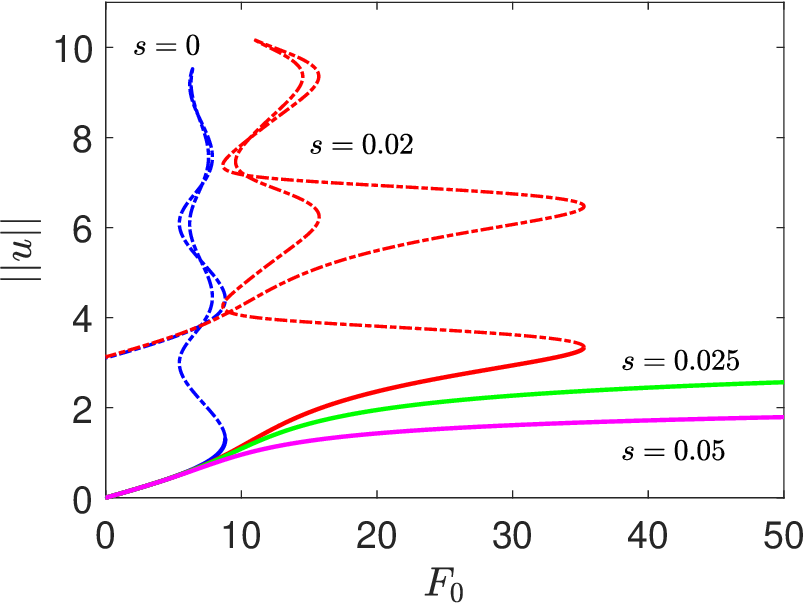}\label{subfig:bifur_F0_vs_norm_rm_0_5_m_5}}\quad
		\subfloat[Saturable, $r_{\min}=0.5$]{\includegraphics[scale=.452]{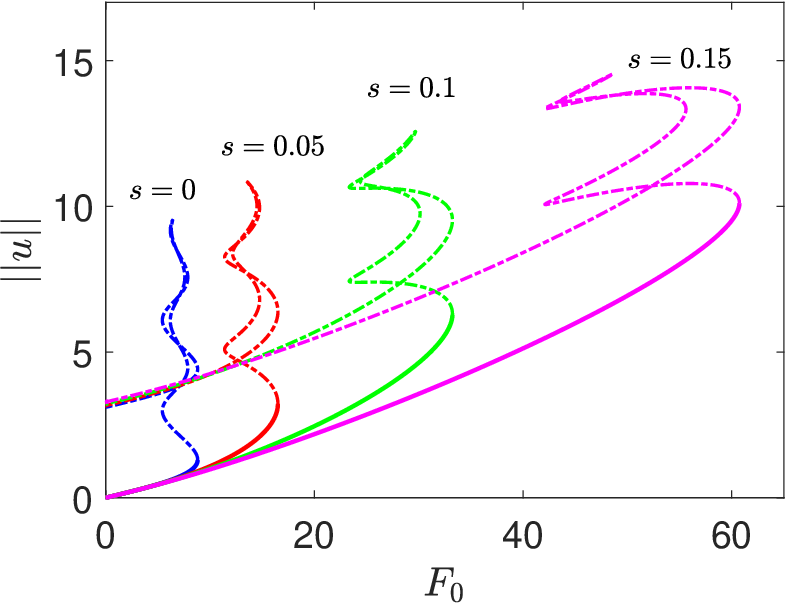}\label{subfig:bifur_F0_vs_norm_satur_rm_0_5_m_5}}\\
		\subfloat[Cubic--quintic, $r_{\min}=2$]{\includegraphics[scale=.452]{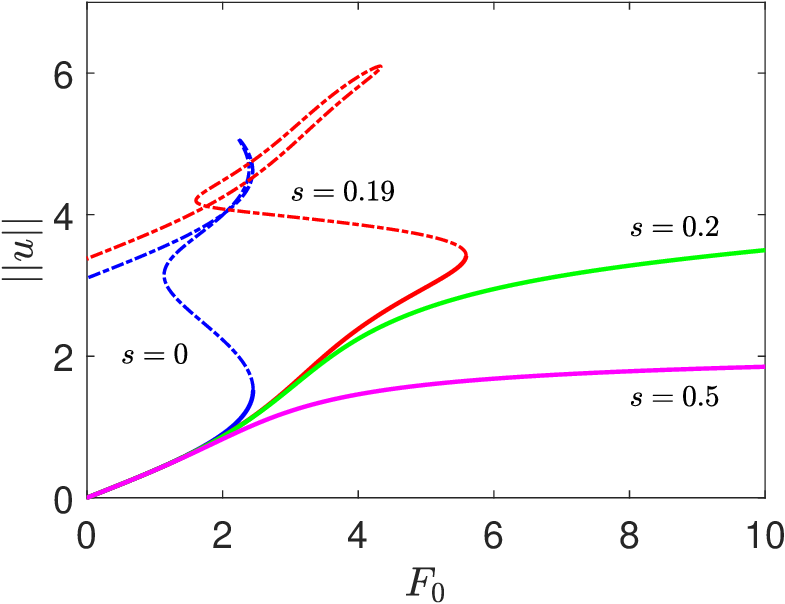}\label{subfig:bifur_F0_vs_norm_rm_2_m_5}}\quad
		\subfloat[Saturable, $r_{\min}=2$]{\includegraphics[scale=.452]{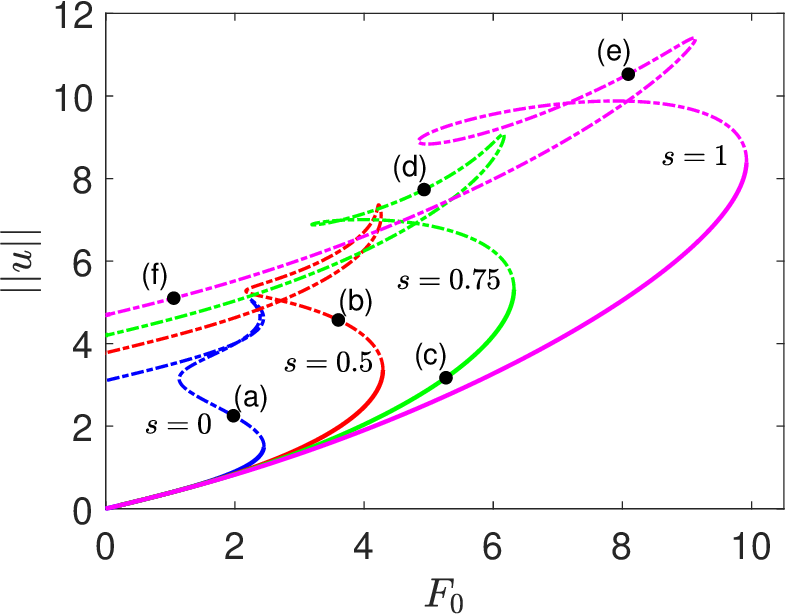}\label{subfig:bifur_F0_vs_norm_satur_rm_2_m_5}}
		\caption{
			Bifurcation diagrams for $m=5$ comparing the cubic--quintic and
			saturable nonlinearities at $r_{\min}=0.5$ and $r_{\min}=2$ with different values of $s$. 
		}
		\label{fig:bifur_F0_vs_norm_rm_2_m_5}
	\end{figure*}
	
	{Figure~\ref{fig:bifur_F0_vs_norm_rm_2_m_5} compares the bifurcation
		structure for $m=5$ in the cubic--quintic and saturable models.
		In the cubic--quintic case, the branches display sharp turning points
		and a pronounced dependence on the driving amplitude $F_0$, reflecting
		the competition between the focusing cubic and defocusing quintic terms.
		The critical value of the drive, denoted by $F_{0C}$, corresponds to the
		turning point (saddle--node) where the stable and unstable branches meet.
		As the quintic coefficient $s$ increases, this turning point shifts to
		larger $F_0$ and eventually disappears, so that the branch becomes
		monotonic and no finite $F_{0C}$ can be defined.
		Physically, this occurs when the defocusing contribution dominates,
		preventing the formation of localized states strong enough to trigger
		supratransmission.}
	
	{
		In contrast, the saturable model shows smoother growth of the norm and
		retains a clear saddle--node structure for all examined values of $s$.
		Because the effective nonlinearity in the saturable model does not become
		defocusing at large amplitudes but instead saturates to a finite value,
		a turning point always exists, ensuring a well-defined critical
		amplitude. The comparison illustrates that, while both nonlinearities
		regularize high intensities, only the saturable response maintains a
		persistent supratransmission threshold as $s$ increases.		
	}
	
	\begin{figure*}[thbp!]
		\centering
		\subfloat[]{\includegraphics[scale=.4]{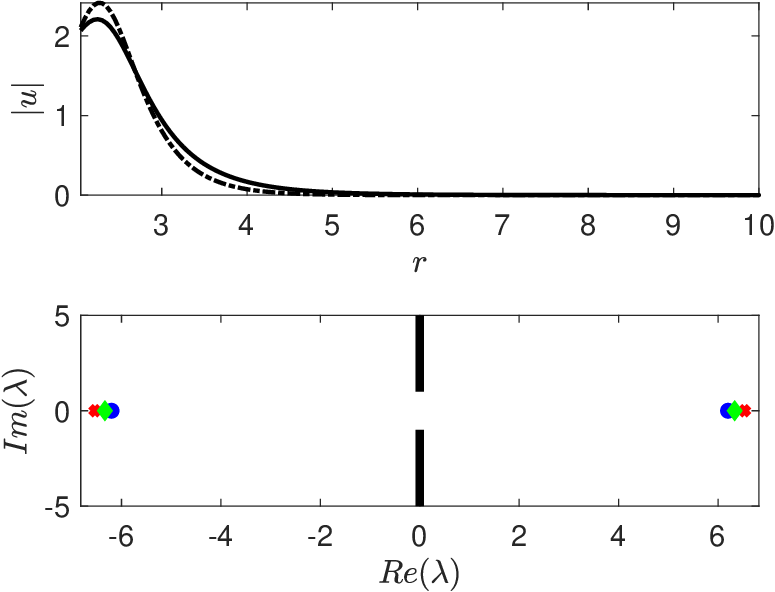}\label{subfig:prof_satur_a}}\hspace{0.5em}
		\subfloat[]{\includegraphics[scale=.4]{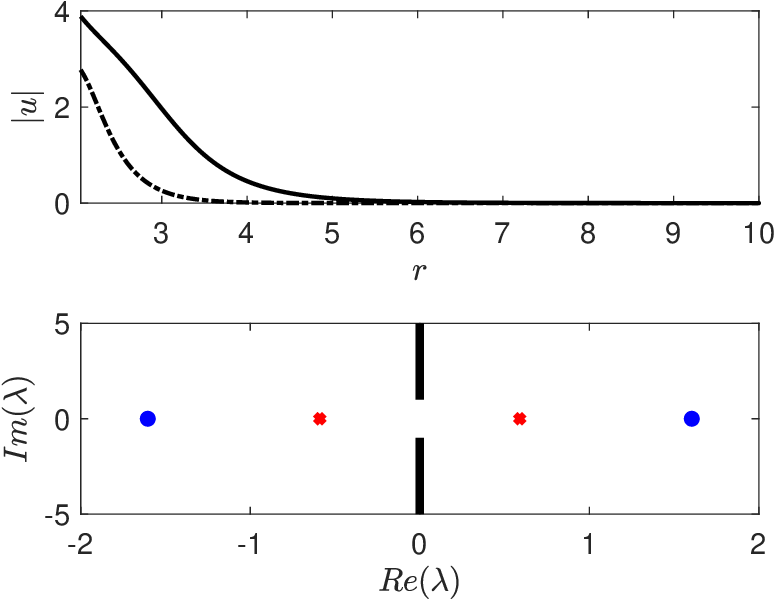}\label{subfig:prof_satur_b}}\hspace{0.5em}
		\subfloat[]{\includegraphics[scale=.4]{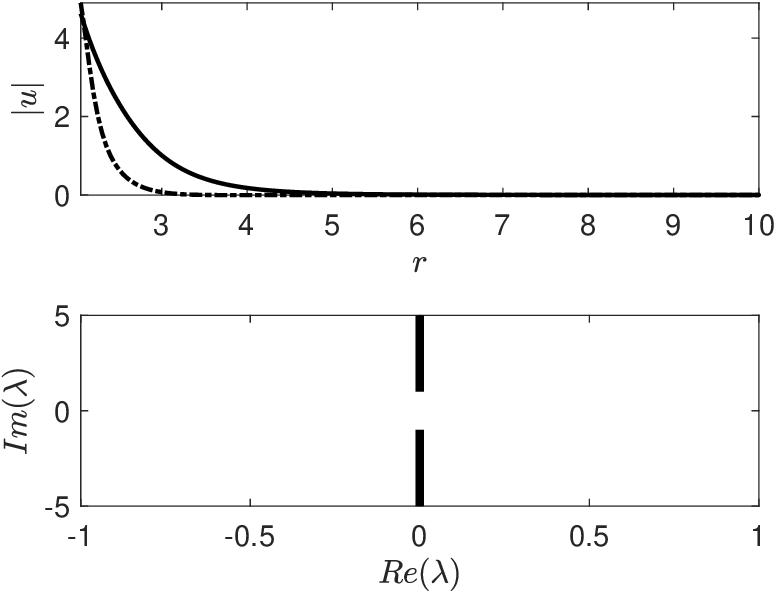}\label{subfig:prof_satur_c}}\\
		\subfloat[]{\includegraphics[scale=.4]{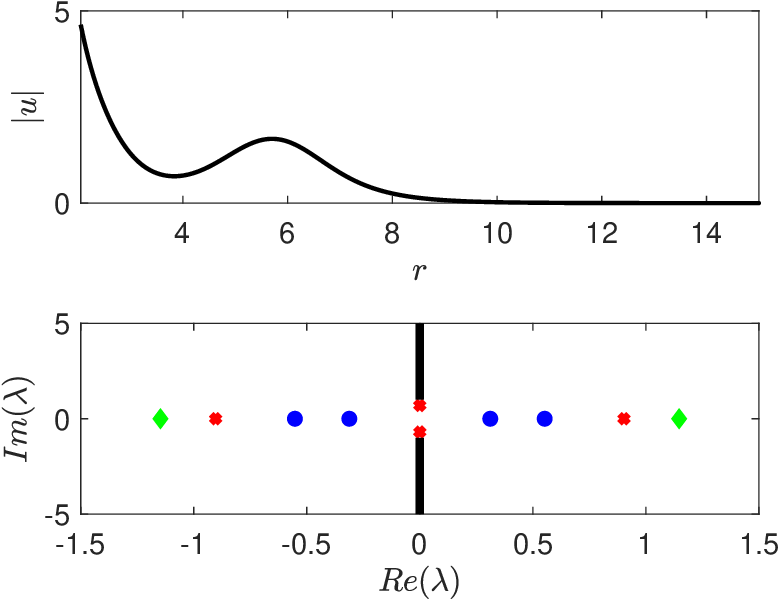}\label{subfig:prof_satur_d}}\hspace{0.5em}
		\subfloat[]{\includegraphics[scale=.4]{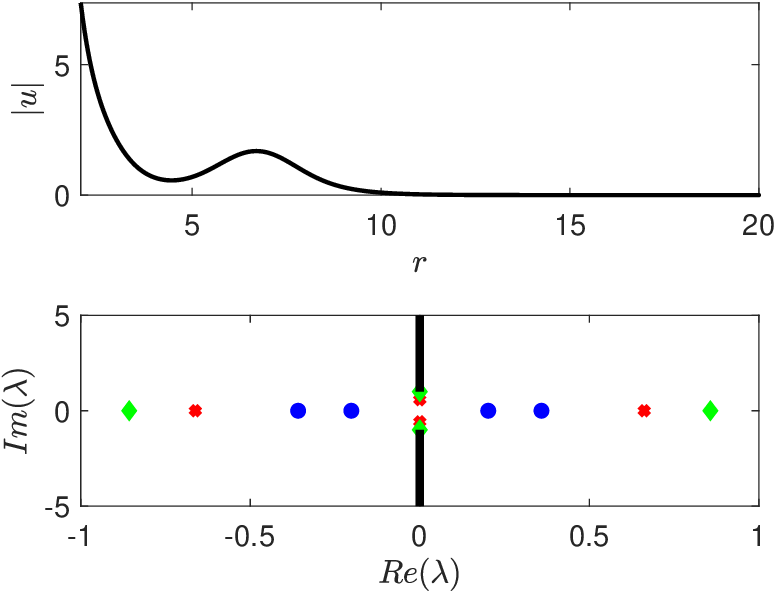}\label{subfig:prof_satur_e}}\hspace{0.5em}
		\subfloat[]{\includegraphics[scale=.4]{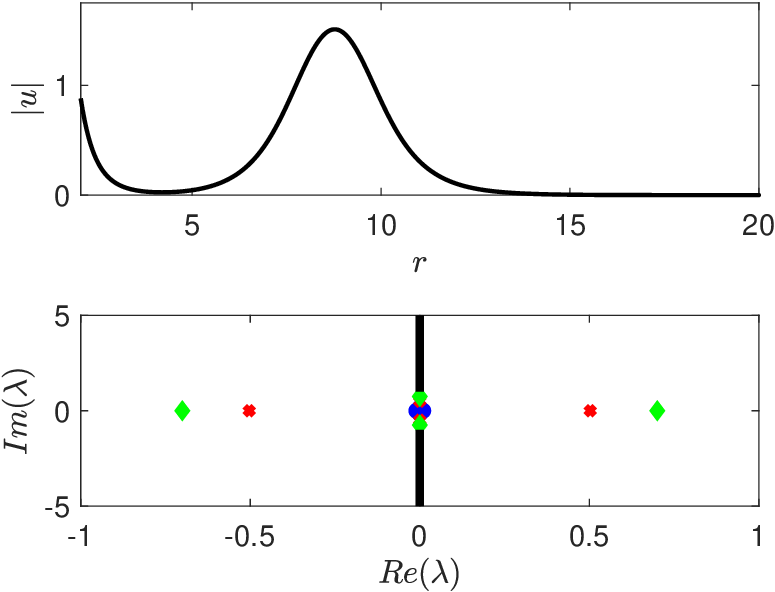}\label{subfig:prof_satur_f}}
		\caption{
			Same as Fig.~\ref{fig:prof_bifur}, but for the bifurcation diagram shown in Fig.~\ref{subfig:bifur_F0_vs_norm_satur_rm_2_m_5}, corresponding to the saturable nonlinearity with $r_{\min}=2$.
		}
		\label{fig:prof_bifur_satur}
	\end{figure*}
	
	
	{Figure~\ref{fig:prof_bifur_satur} shows representative standing-wave
		profiles and corresponding spectra for the saturable nonlinearity at
		$r_{\min} = 2$, sampled from the bifurcation diagram in
		Fig.~\ref{subfig:bifur_F0_vs_norm_satur_rm_2_m_5}.
		Each panel (a)-(f) corresponds to a different value of the saturation
		parameter $s$ and to specific points along the branches.
		For $s = 0$ [panel~(a)], the solution corresponds to the cubic case and
		is unstable.
		For $s = 0.5$ [panel~(b)], saturation weakens the effective nonlinearity,
		but the mode remains unstable.
		At $s = 0.75$, the lower-norm state [panel~(c)] is spectrally stable with
		eigenvalues on the imaginary axis, while the higher-norm state on the
		same branch [panel~(d)] becomes unstable, indicating a transition near
		the turning point.
		For $s = 1$, both the high-amplitude [panel~(e)] and low-amplitude
		[panel~(f)] states are unstable, characterized by complex eigenvalue
		pairs associated with azimuthal perturbations.		
		The clustering of eigenvalues near the imaginary axis for the more
		saturated cases illustrates how saturation limits the growth rate of
		unstable modes.
		Nonetheless, azimuthal modes with $q \ge 1$ can still destabilize certain
		profiles, indicating that saturation regularizes but does not completely
		eliminate multidimensional instabilities.
		Overall, increasing $s$ expands the range of stable standing waves at
		intermediate amplitudes, yet high driving levels eventually reintroduce instability on the upper branches.		
	}
	
	\begin{figure*}[thbp!]
		\centering
		\subfloat[Cubic ($s=0$)]{\includegraphics[scale=.42]{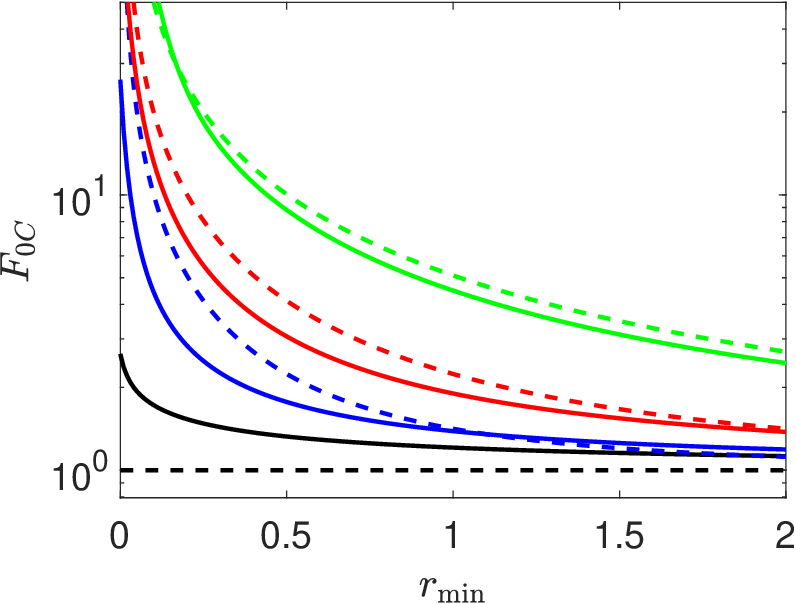}\label{subfig:F0C_vs_rm_s_0}}\quad
		\subfloat[Cubic-quintic ($s=0.02$)]{\includegraphics[scale=.42]{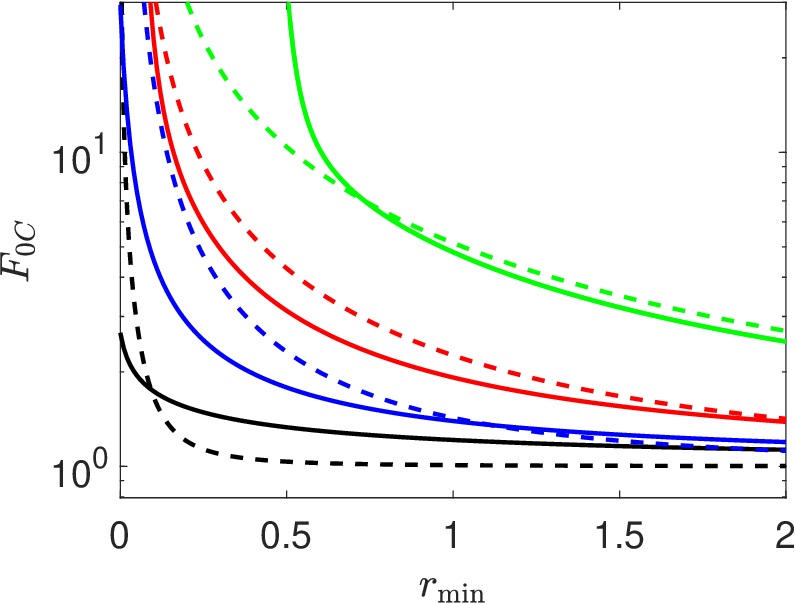}\label{subfig:F0C_vs_rm_s_0_02}}
		\subfloat[Saturable ($s=0.02$)]{\includegraphics[scale=.42]{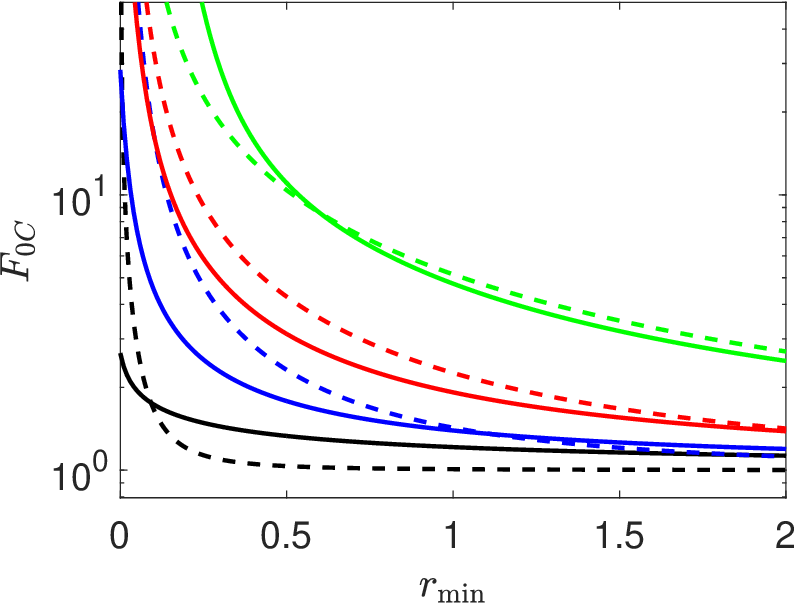}\label{subfig:F0C_vs_rm_satur_s_0_02}}
		\caption{
			{Critical drive amplitudes $F_{0C}$ at the first turning points
				(saddle-node bifurcations) of the standing-wave branches are shown as functions
				of the inner radius $r_{\min}$ for the cubic, cubic--quintic, and saturable
				nonlinearities.  
				Curves represent azimuthal indices
				$m=0$ (black), $m=1$ (blue), $m=2$ (red), and $m=5$ (green).  
				Solid lines are numerical results, and dashed lines are the variational
				approximation~\eqref{A}.  
				In all cases, $F_{0C}$ grows as $r_{\min}$ decreases or as $m$ increases,
				showing that stronger confinement or higher azimuthal charge requires a larger
				driving amplitude to initiate supratransmission.
			}
		}
		\label{fig:F0C_vs_rm}
	\end{figure*}
	
	{Figure~\ref{fig:F0C_vs_rm} compares the critical driving amplitude
		$F_{0C}$ at the first turning point of the standing-wave branches for the
		cubic, cubic--quintic, and saturable nonlinearities.
		In all cases, $F_{0C}$ increases sharply as the inner radius $r_{\min}$
		decreases, showing that stronger confinement requires a larger drive to
		initiate supratransmission.
		For the cubic model [panel~(a)], the dependence on $r_{\min}$ and azimuthal charge $m$ is nearly monotonic, with higher $m$ requiring larger thresholds because the azimuthal phase circulation produces an effective radial repulsion that resists localization near the inner boundary and produces localized fields that are broader and more weakly peaked radial profiles~\cite{kivshar2003optical}.	The cubic--quintic case [panel~(b)] retains this trend but displays larger
		overall values of $F_{0C}$, reflecting the additional defocusing
		contribution of the quintic term, which counteracts self-focusing and
		raises the onset of transmission.
		The saturable model [panel~(c)] exhibits a similar qualitative behavior but
		with slightly lower thresholds at large $r_{\min}$, consistent with the bounded
		effective nonlinearity that limits the growth of the field at high
		amplitudes.	
	}
	
	{
		In the figure, we also show the variational approximation~\eqref{A} (dashed lines), which will be derived in the following section. The approximation reproduces the numerical results well, especially for small azimuthal indices and larger 	$r_{\min}$, where curvature effects are weaker. For higher $m$, deviations increase because the simplified radial ansatz
		underestimates the influence of azimuthal modulation.
		Overall, both the numerical and analytical results confirm that geometric
		confinement and topological charge play key roles in determining the
		critical amplitude for supratransmission in two-dimensional settings.
	}
	\subsection{Variational approximation}\label{Sec:VA}
	
	Supratransmission has been known mostly due to the drive that exceeds the corresponding solitary wave amplitude of the problem in the \emph{infinite} domain \cite{geniet2002energy,geniet2003nonlinear,khomeriki2004nonlinearband,khomeriki2004nonlinear} (even though it is not always the case, see \cite{susanto2023surge}). However, explicit expressions of localized solutions for the two-dimensional NLS are unavailable. Consequently, we require tractable approximate solutions. In the following, we adopt a variational approach. 
	
	For the analysis, we will consider the cubic--quintic case. 
	The governing equation \eqref{main1} \emph{without} the inner ring has the action functional $\mathcal{S}$ and the Lagrangian $\mathcal{L}$ \cite{afanasjev1995rotating,caplan2009azimuthal,caplan2012existence}
	\begin{equation}
		\mathcal{S}=\int_{t_0}^{t_1} \mathcal{L}\, dt,\,\qquad \mathcal{L} = \int_0^{2\pi}\int_0^\infty L\, r\, dr\, d\theta,
	\end{equation}
	where the Lagrangian density $L$ is given by 
	\begin{equation}
		L = \frac{i}{2}\left(u^*u_t-u u_t^*\right) +
		\left|u_r\right|^2 + \frac{1}{r^2}\left|u_\theta\right|^2
		- |u|^4+\frac{s}3|u|^6.
		\label{lagden}
	\end{equation}
	Equation \eqref{main1} is obtained from the action function $\mathcal{S}$ by evaluating its variational derivative 
	\begin{equation}
		\frac{\delta\mathcal{S}}{\delta u^*} = \frac{\partial}{\partial t} \frac{\partial L}{\partial u_t^*} + \frac{\partial}{\partial r} \frac{\partial L}{\partial u_r^*} + \frac{\partial}{\partial \theta} \frac{\partial L}{\partial u_\theta^*} - \frac{\partial L}{\partial u^*} = 0.
	\end{equation}
	
	To perform the variational approximation method, we insert the following vortex ansatz following \cite{afanasjev1995rotating,caplan2009azimuthal,caplan2012existence}
	\begin{equation}
		\label{var_sep}
		u(r,\theta,t)= f(r)e^{i(m\theta-\Omega t)},
	\end{equation}
	where we use a one-dimensional soliton sech ansatz for the radial profile
	\begin{equation}
		\label{sech_ansatz}
		f(r) =  A\sech\left( A(r-r_c)\right),
	\end{equation}
	with parameters $A$ and $r_c$ corresponding to the amplitude and center of the ring, respectively. Substituting the ansatz into the Lagrangian, we obtain 
	\begin{equation}
		\label{var_langC}
		\mathcal{L} = 2\pi\left(\Omega \, C_1 + C_2 + m^2\, C_3 - C_4 + \frac s3C_5\right),
	\end{equation}
	where, writing $E=e^{2Ar_c}$ and assuming $r_c$ to be large, the constants are given by 
	\begin{widetext}
			\begin{subequations}
				\label{constants}
				\begin{align}
					C_1&= \int_0^\infty \left|f(r)\right|^2r\,dr = \ln(E+1) \approx 2Ar_c,\\  
					C_2&= \int_0^\infty \left|\frac{df}{dr}\right|^2r\,dr = \frac13\,{\frac {{A}^{2} \left[ \left({E}^{2}+2\,E+1\right)\ln  \left( E+1 \right) +2\,E  \right] }{{E}^{2}+2
							\,E+1}}\approx \frac23 A^3r_c,\\
					C_3&=\int_0^\infty\!\frac{1}{r^2}\left|f(r)\right|^2r\,dr \approx \frac{1}{r_c}\int_0^\infty\!\left|f(r)\right|^2\,dr=\frac1{r_c}\frac{2EA}{E+1}\approx\frac{2A}{r_c},\\ 
					C_4&=\int_0^\infty\!\left|f(r)\right|^4r\,dr = \frac23\,{\frac {{A}^{2} \left[ \left({E}^{2}+2\,E+1\right)\ln  \left( E+1 \right) -E  \right] }{{E}^{2}+2\,E
							+1}} \approx \frac43 A^3r_c,\\
					C_5&=\int_0^\infty\!\left|f(r)\right|^6r\,dr = \frac4{15}{\frac {{A}^{4} \left[ \left(2\,{E}^{4}+8\,{E}^{3}+12\,{E}^{2}+8\,E+2\right)\ln  \left( E+1 \right) -2\,{E}^{3} 
							-7\,{E}^{2}  -2\,E  \right] }{{E}^{4}+60\,{E}^{3}+90\,{E}^{2}+60\,E+15}} \approx \frac{16}{15}A^5r_c.
				\end{align}
			\end{subequations}
		\end{widetext}
		Note that the $C_3$ integral does not converge due to the singularity at $r=0$ and following \cite{afanasjev1995rotating,caplan2009azimuthal}, we take the approximation $r\approx r_c$. 
		
		Note that the amplitude $A$ is unknown and needs to be optimized. The Euler-Lagrange equation \[{\partial \mathcal{L}}/{\partial A}=2\,{r_c}\,{\Omega}-2\,{A}^{2}{r_c}+\frac{2 m^2}{{r_c}}+{
			\frac{16}{9} {s{A}^{4}{r_c}}}
		= 0\] yields
		\begin{equation}
			\begin{array}{rcl}
				A^2 &=&\displaystyle \frac3{16}{\frac {3\,r_c-\sqrt {-32\,{\Omega}\,s{r_c}^{2}-32\,m^2s
							+9\,{r_c}^{2}}}{sr_c}}\\
				&\approx&\displaystyle \frac{m^2}{r_c^2}+\Omega+{\frac {8\, \left({\Omega} {r_c}^{2}+m^2 \right)^{2}s}{9\,r_c^{4}}}+\dots.
			\end{array}
			\label{A}
		\end{equation}
		Taking $r_c=r_{\min}$, the amplitude \eqref{A} will approximate the critical drive. Equations \eqref{var_sep}, \eqref{sech_ansatz}, \eqref{A} also approximate solutions along the lower branch of the bifurcation diagram up to near the turning point. However, the vortex center $r_c$ is now determined from the equation $f(r_{\min})=F_0$, i.e., 
		\begin{equation}
			r_c = r_{\min} \pm \frac{1}{A}\sech^{-1}(F_0/A).
			\label{rc}
		\end{equation}
		{We plot our approximate solutions in
			Figs.~\ref{subfig:prof_a}--\ref{subfig:prof_c} and
			\ref{subfig:prof_satur_a}--\ref{subfig:prof_satur_c}.
			For stable solutions on the lower branch, the negative sign in
			Eq.~\eqref{rc} is used.
			For unstable solutions near or beyond the turning point, the same
			relation~\eqref{rc} is applied with the positive sign, corresponding to
			profiles whose centers are shifted outward from the inner boundary.
			In both cases, the variational profiles show good agreement with the
			numerical solutions.
		}
		
		The same variational approximation can be carried out for the case of saturable nonlinearity. However, we leave this case to the interested reader as the calculation is cumbersome \cite{syafwan2012variational}, despite similarity. For the case of cubic--quintic nonlinearity, we can explain the presence of a critical $s$ above which there is no critical threshold drive amplitude as follows. 
		
		Consider the case when $r_{\min}\to\infty$, in which case the time-independent equation \eqref{main2} becomes the standard one-dimensional nonlinear Schr\"odinger equation 
		\begin{equation}
			U_{rr}-\Omega U + 2U^3-sU^5 =0.
			\label{temp}
		\end{equation}
		The invariant manifold from the equilibrium $U=0$ is given by 
		\begin{equation}
			U_r^2 = \Omega U^2 -U^4+\frac{s}{3} U^6.
		\end{equation}
		The curve $U_r^2$ has zeros at $0,\sqrt{3\pm\sqrt{9-12s\Omega}}/\sqrt{2s}$ 
		(and their negative counterparts). The last two zeros merge at $s_m={3\Omega}/4$ and then disappear for $s$ above the value. This is a global bifurcation point, where a homoclinic trajectory from and to $(U,U_r)=(0,0)$ breaks and becomes an unbounded manifold. This explains the numerical result that for $s>3\Omega/4$, there is no critical drive amplitude, which physically corresponds to the situation where the defocusing nonlinearity dominates the focusing one. This global change does not occur with saturable nonlinearity, implying that there will always be a critical drive amplitude. Figure~\ref{fig:bifur_F0_vs_norm_rm_2_m_5} further suggests that the critical value $s_m$ at which the supratransmission threshold disappears decreases as $r_{\min}$ becomes smaller, indicating that geometric confinement enhances the effective defocusing contribution.
		
		\section{Nonlinear dynamics of supratransmission}\label{sec:dynamics}
		
		{ 
			After discussing the standing-wave solutions, we turn to the
			time-dependent governing equations \eqref{main1}--\eqref{drive}.
			We integrate them numerically using the
			fourth-order Runge--Kutta method. Radial and azimuthal derivatives are
			approximated by central finite differences, using two-point stencils for
			first derivatives and three-point stencils for second derivatives
			\cite{strikwerda2004finite}.
			The computational domain is truncated at $r=r_{\max}$, and an absorbing
			layer is introduced by adding the term $-i\alpha(r)u$ to the right-hand
			side of Eq.~\eqref{main1}. The absorption profile is taken as
			\[
			\alpha(r)=
			\begin{cases}
				0, & r < r_b, \\[4pt]
				\exp\!\bigl(r-r_b\bigr)-1, & r \ge r_b,
			\end{cases}
			\]
			where $r_b<r_{\max}$ marks the onset of the absorbing region.
			Thus, $\alpha(r)$ vanishes in the interior and increases smoothly for
			$r\ge r_b$, reaching its largest values near $r_{\max}$.
			This damping layer reduces spurious reflections from the outer boundary. In this section, we set $\tau = 50$ in the adiabatic ramp defined in Eq.~\eqref{eq:inc}.
		}
		
		
		
		\begin{figure*}[tbhp!]
			\centering
			\subfloat[$t=0.2$]{\includegraphics[scale=.45]{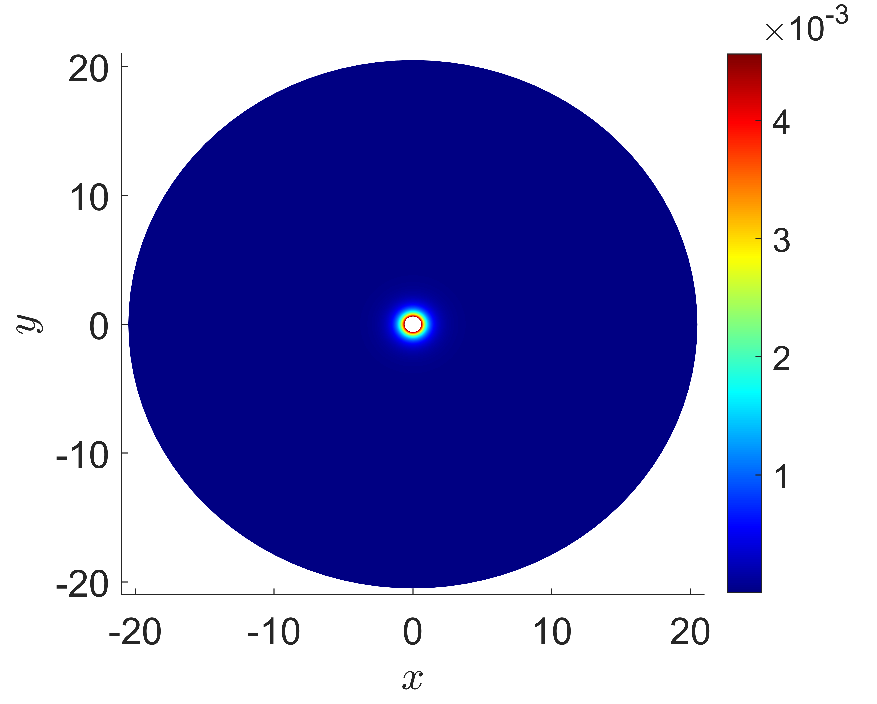}\label{subfig:TD_CQ_F0=1_5_s=0_02_m=0_q=0_rm=0_5_t=0_2}}
			\subfloat[$t=116$]{\includegraphics[scale=.45]{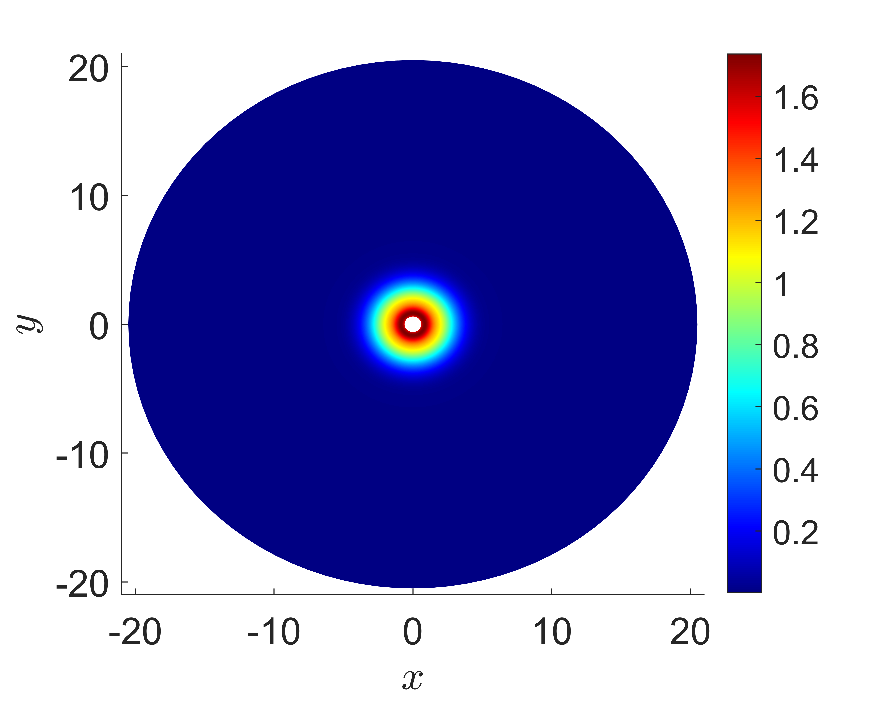}\label{subfig:TD_CQ_F0=1_5_s=0_02_m=0_q=0_rm=0_5_t=116}}\\
			\subfloat[$t=117$]{\includegraphics[scale=.45]{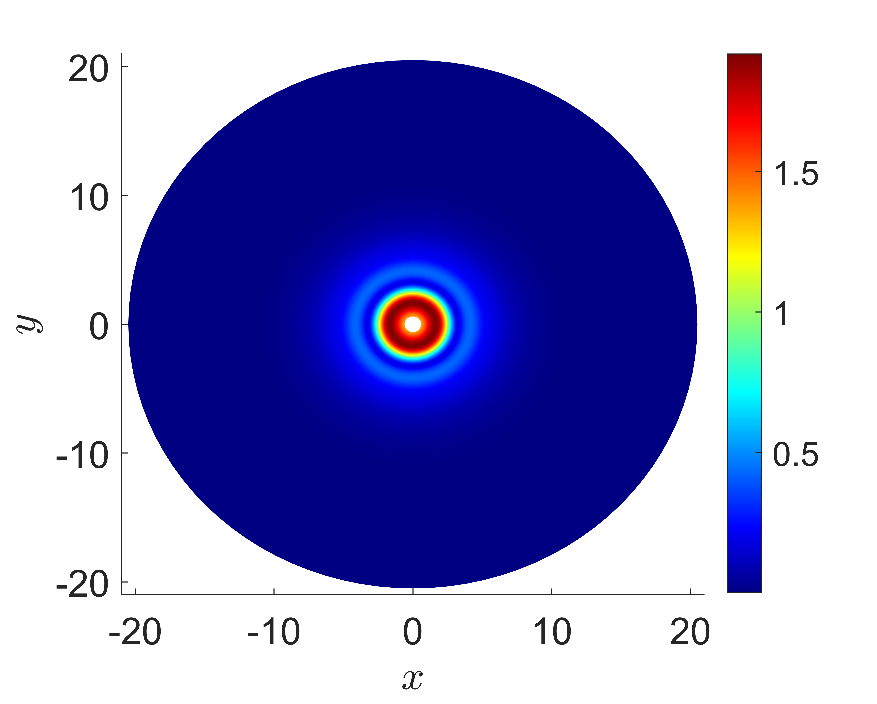}\label{subfig:TD_CQ_F0=1_5_s=0_02_m=0_q=0_rm=0_5_t=117}}
			\subfloat[$t=120$]{\includegraphics[scale=.45]{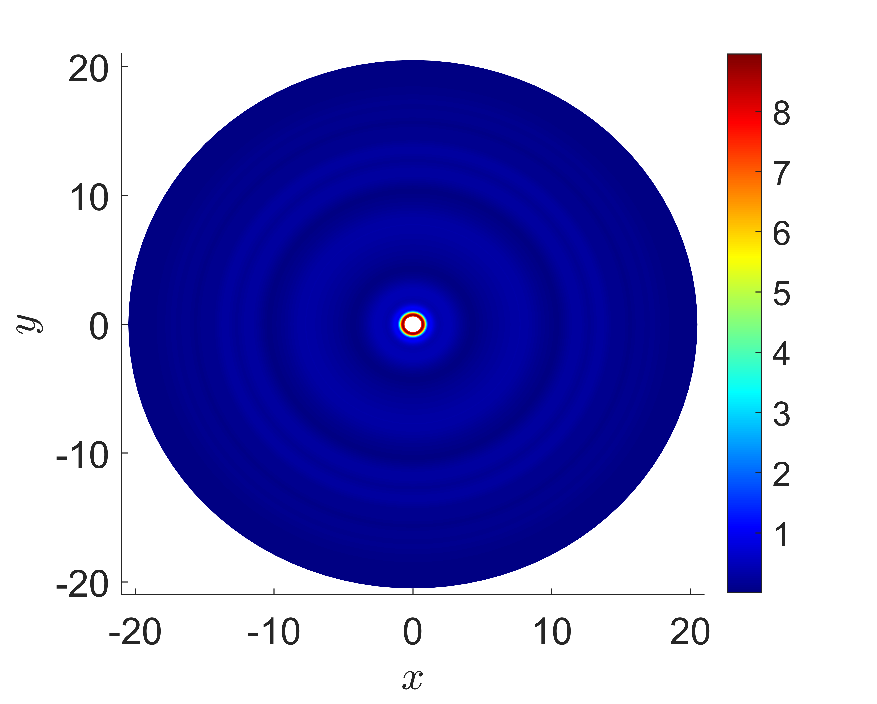}\label{subfig:TD_CQ_F0=1_5_s=0_02_m=0_q=0_rm=0_5_t=120}}\\
			\subfloat[$t=121$]{\includegraphics[scale=.45]{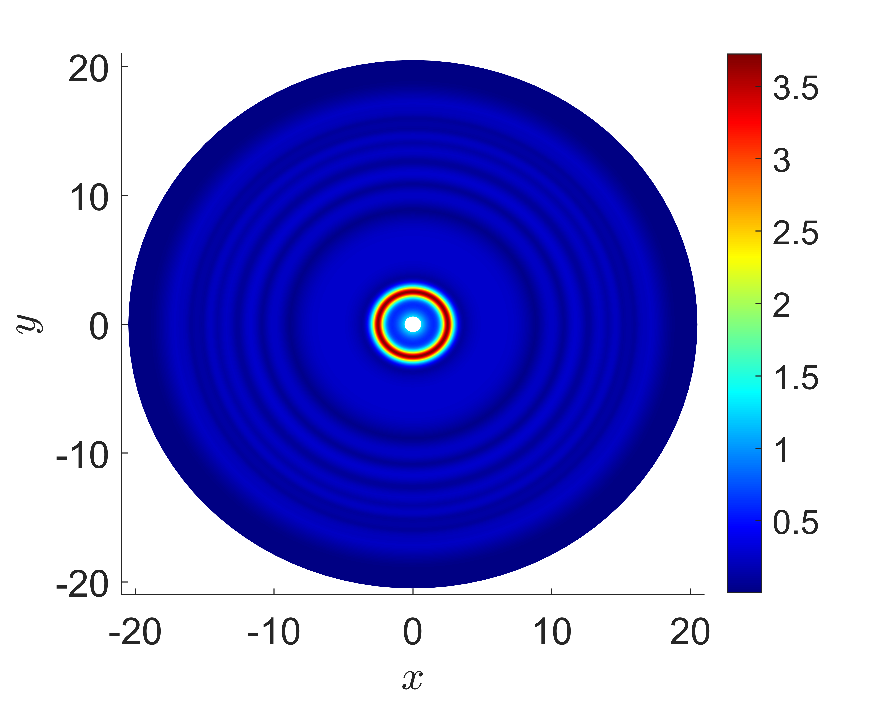}\label{subfig:TD_CQ_F0=1_5_s=0_02_m=0_q=0_rm=0_5_t=121}}
			\subfloat[$t=123$]{\includegraphics[scale=.45]{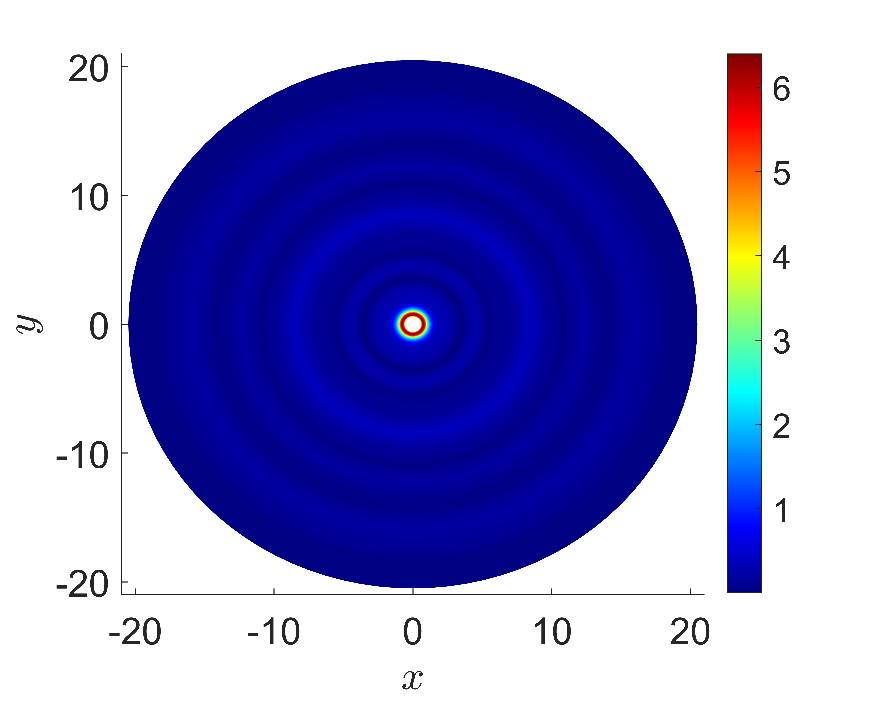}\label{subfig:TD_CQ_F0=1_5_s=0_02_m=0_q=0_rm=0_5_t=123}}
			\caption{
				Time evolution of the cubic--quintic system for
				$F_0=1.5$, $r_{\min}=0.5$, $s=0.02$, and $m=0$. Panels (a)--(f) show
				snapshots of the profile $|u|$ at selected times, illustrating the transition from the initial evanescent response to the emergence and recurrence of a radially propagating pulse. }
			\label{fig:TD_CQ_F0=10_s=0_02_m=0_q=0_rm=0_5}
		\end{figure*}
		
		%
		
		{Figure~\ref{fig:TD_CQ_F0=10_s=0_02_m=0_q=0_rm=0_5} shows the time
			evolution of the cubic--quintic system with the zero initial condition.
			The parameters are $F_0 = 1.5$, $r_{\min} = 0.5$, $s = 0.02$, $m = 0$, corresponding to a regime slightly above the turning point of
			the standing-wave branch. Panels (a)-(f) present snapshots of the field
			amplitude $|u(x,y,t)|$ at selected times.}		
		
		{
			At early times [panel~(a), $t = 0.2$], the system responds linearly to the
			drive, producing a weak evanescent field near the inner boundary. As the
			driving amplitude gradually reaches its full strength, a nonlinear
			transition occurs: a localized bright ring begins to form and separate from
			the boundary [panels~(b)-(c), $t = 116$-$117$]. The pulse then propagates
			outward while maintaining radial symmetry. Because the input
			energy is limited, the outward motion slows down and the pulse starts to
			move back toward the center [panel~(d), $t = 120$]. This
			expansion-contraction sequence repeats periodically, as shown in
			panels~(e)-(f), generating a recurrent radial oscillation of a
			ring-shaped solitary structure.
			These results indicate that once the driving amplitude exceeds the
			supratransmission threshold, energy is transmitted through the formation of
			a self-sustained, radially propagating pulse. The resulting oscillatory motion reflects the balance between the robustness and stability of the soliton ring and the limited available energy, which prevents further outward expansion and instead produces a sloshing motion near the inner region.
		}
		
		\begin{figure*}[thbp!]
			\centering
			{\includegraphics[scale=.7]{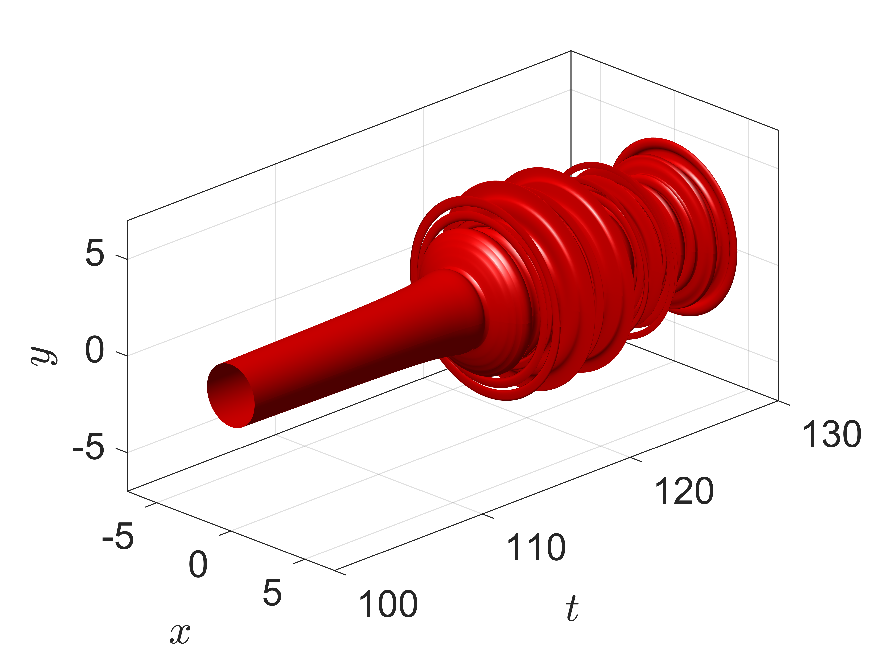}}
			\caption{Isosurface visualization of the time evolution corresponding to Fig.\ \ref{fig:TD_CQ_F0=10_s=0_02_m=0_q=0_rm=0_5}. The isosurface is plotted for the solution level set $|u| = 0.4$, illustrating the spatiotemporal dynamics and the transition from evanescent to recurrent dynamics of expanding and contracting ring solitons.}
			\label{subfig:isosurface_CQ_F0=1_5_s=0_02_m=0_rm=0_5}
		\end{figure*}
		
		Figure~\ref{subfig:isosurface_CQ_F0=1_5_s=0_02_m=0_rm=0_5} shows an
		isosurface plot of the cubic--quintic evolution corresponding to
		Fig.~\ref{fig:TD_CQ_F0=10_s=0_02_m=0_q=0_rm=0_5}. The surface
		$|u|=0.4$ is rendered in $(r,\theta,t)$ to highlight the
		spatiotemporal motion of the emitted pulse. The plot reveals a sequence
		of outward and inward excursions of a ring-shaped structure, consistent
		with the recurrent radial dynamics observed in the snapshots. The
		isosurface representation makes the periodic expansion--contraction cycle
		clear and provides a compact view of the temporal dynamics.
		
		\begin{figure*}[thbp!]
			\centering
			\subfloat[$t=0.2$]{\includegraphics[scale=.45]{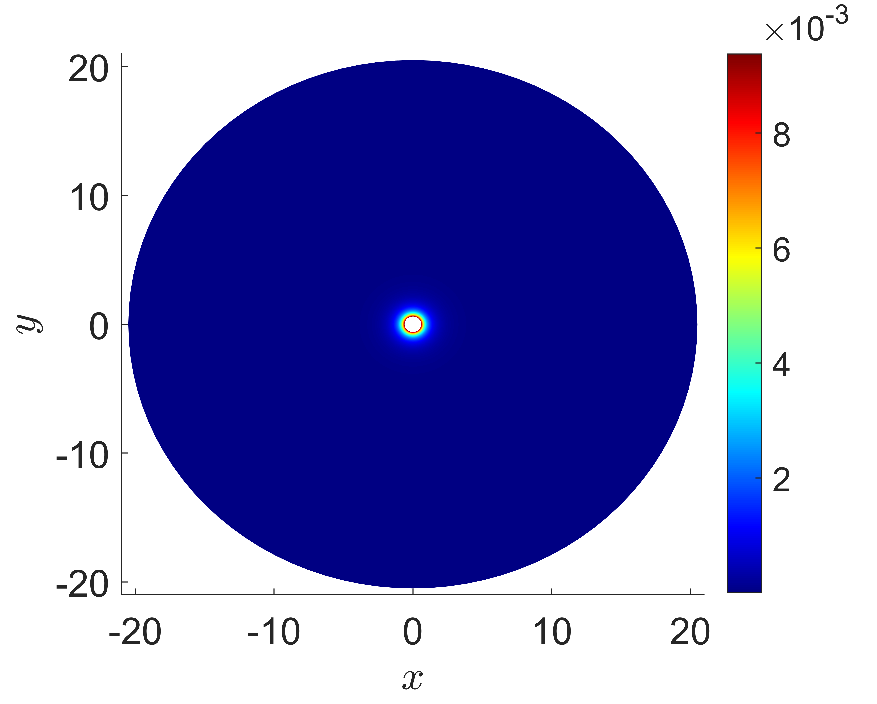}\label{subfig:TD_CQ_F0=3_5_s=0_02_m=2_q=0_rm=0_5_t=0_2}}
			\subfloat[$t=116$]{\includegraphics[scale=.45]{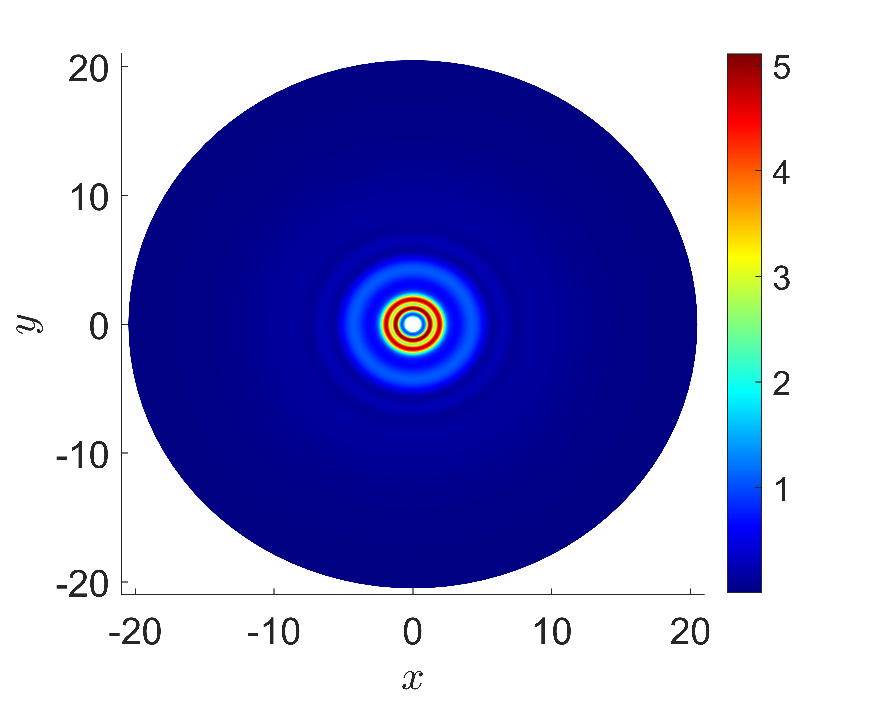}\label{subfig:TD_CQ_F0=3_5_s=0_02_m=2_q=0_rm=0_5_t=116}}\\
			\subfloat[$t=117.2$]{\includegraphics[scale=.45]{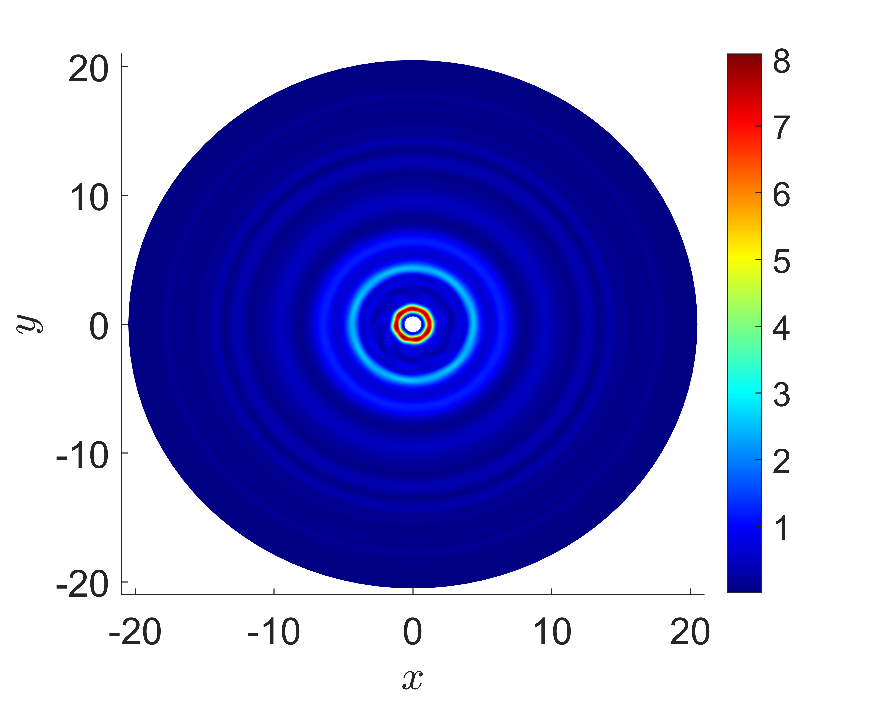}\label{subfig:TD_CQ_F0=3_5_s=0_02_m=2_q=0_rm=0_5_t=117_2}}
			\subfloat[$t=117.4$]{\includegraphics[scale=.45]{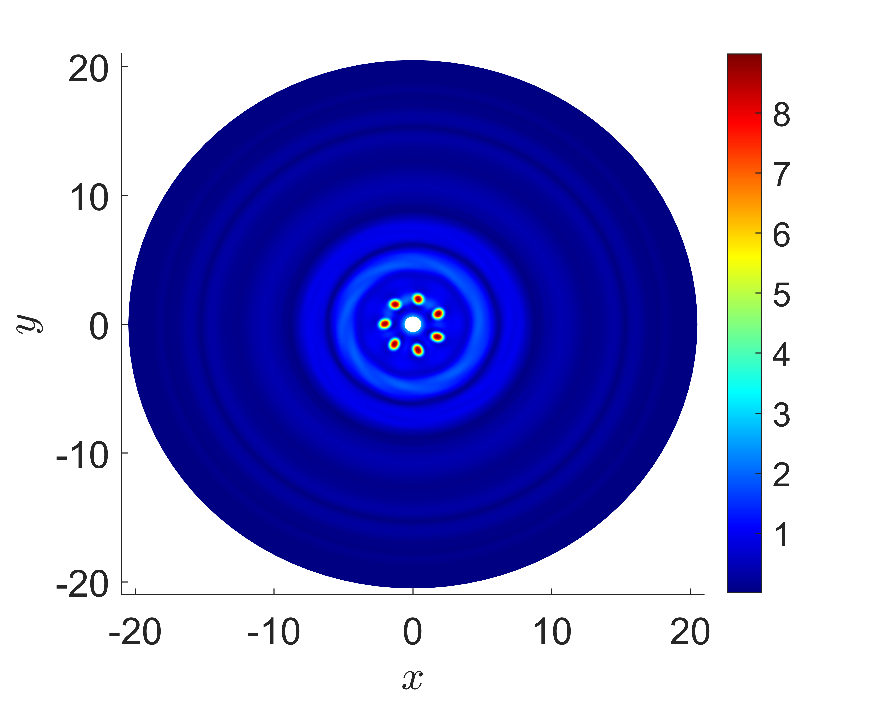}\label{subfig:TD_CQ_F0=3_5_s=0_02_m=2_q=0_rm=0_5_t=117_4}}\\
			\subfloat[$t=118.2$]{\includegraphics[scale=.45]{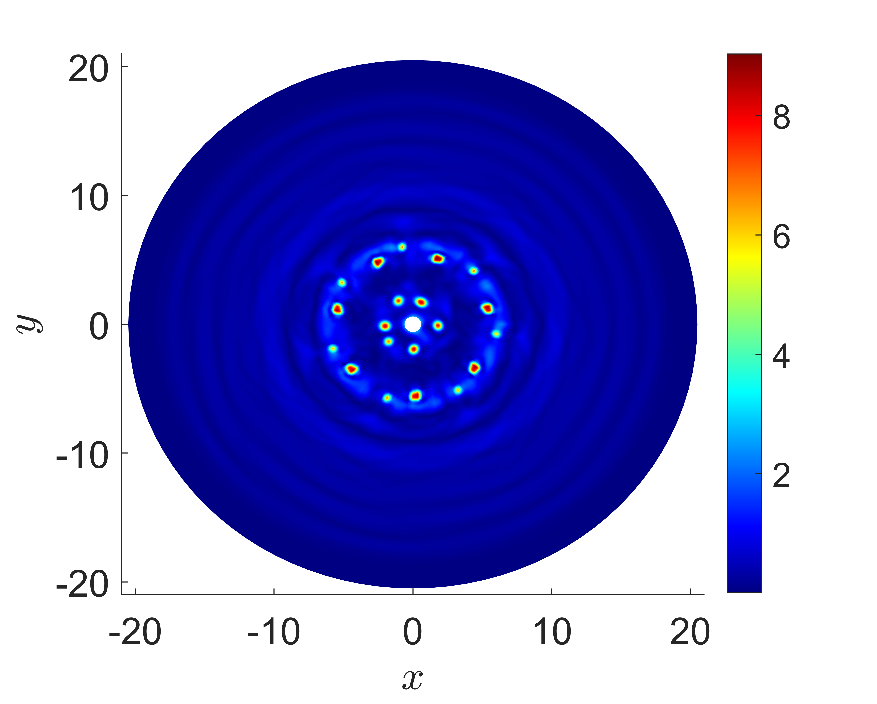}\label{subfig:TD_CQ_F0=3_5_s=0_02_m=2_q=0_rm=0_5_t=118_2}}
			\subfloat[$t=120$]{\includegraphics[scale=.45]{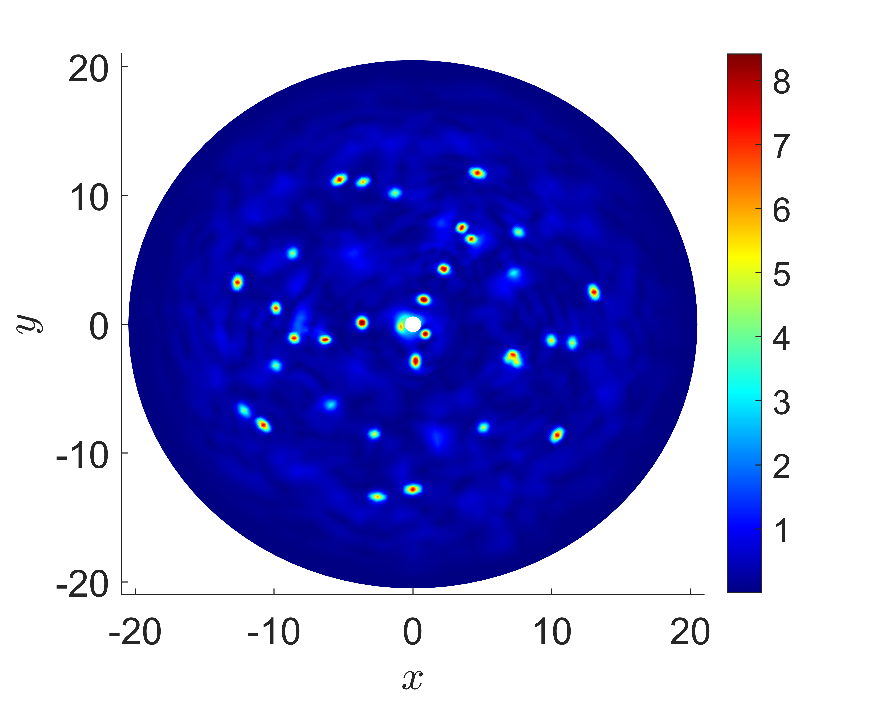}\label{subfig:TD_CQ_F0=3_5_s=0_02_m=2_q=0_rm=0_5_t=120}}
			\caption{
				Time evolution of the cubic--quintic system for $m=2$ with
				$F_0=3.5$, $r_{\min}=0.5$, and $s=0.02$. Panels (a)--(f) show snapshots of
				the wave profile $|u|$ at selected times. The solution remains nearly
				axisymmetric at early times and then develops angular deformation as the
				time progresses, leading to a loss of symmetry and increasingly complex dynamics. The creation of two-dimensional localized solitons is clearly seen. 
			}
			\label{fig:TD_CQ_F0=10_s=0_02_m=2_q=0_rm=0_5}
		\end{figure*}
		
		Figure~\ref{fig:TD_CQ_F0=10_s=0_02_m=2_q=0_rm=0_5} shows the evolution of
		the cubic--quintic system for azimuthal charge $m=2$ and driving
		amplitude $F_0=3.5$, using the same values of $r_{\min}$ and $s$ as in the
		$m=0$ case of Fig.~\ref{fig:TD_CQ_F0=10_s=0_02_m=0_q=0_rm=0_5}. The
		initial response [panel~(a), $t=0.2$] is nearly axisymmetric. As the
		drive increases, a radially symmetric ring soliton forms
		[panel~(b)]. However, this structure is unstable, and angular
		modulation develops [panel~(c)], leading to the formation of distinct
		two-dimensional localized solitons, i.e., pearls, [panel~(d)]. At later times [panels~(e)-(f)], these pearls detach from the inner boundary and propagate outward, marking a clear
		breakdown of radial symmetry. Relative to the $m=0$ evolution, the
		nonzero azimuthal charge accelerates symmetry loss and generates a
		richer sequence of two-dimensional localized excitations.

		\begin{figure*}[thbp!]
			\centering
			{\includegraphics[scale=.7]{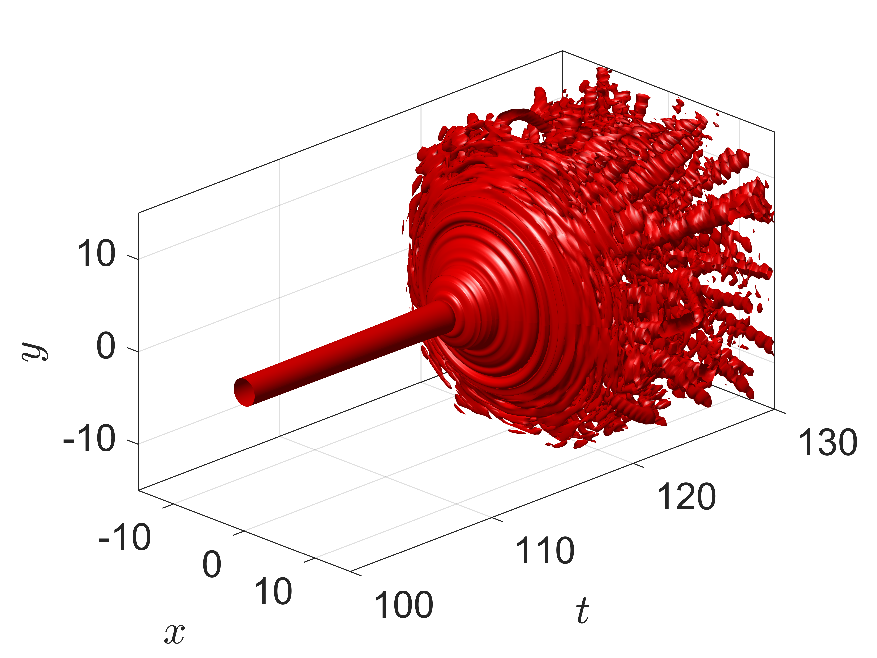}}	
			\caption{Isosurface visualization of the time evolution corresponding to Fig.\ \ref{fig:TD_CQ_F0=10_s=0_02_m=2_q=0_rm=0_5}, for the cubic--quintic nonlinearity with $F_0 = 3.5$, $r_{\min} = 0.5$, $s = 0.02$, and $m = 2$. The isosurface is plotted for the solution height $|u| = 0.4$, illustrating the transition from a stable wave structure to chaotic dynamics as nonlinear instabilities develop.}
			\label{subfig:isosurface_CQ_F0=3_5_s=0_02_m=2_rm=0_5}
		\end{figure*}
		
		Figure~\ref{subfig:isosurface_CQ_F0=3_5_s=0_02_m=2_rm=0_5} presents an
		isosurface plot corresponding to the evolution in
		Fig.~\ref{fig:TD_CQ_F0=10_s=0_02_m=2_q=0_rm=0_5}, using the level
		$|u|=0.4$ to depict the spatiotemporal structure of the field. At
		early times the isosurface remains nearly axisymmetric, consistent with
		the formation of the initial ring soliton. As time increases, angular
		modulation becomes evident, and the isosurface captures the detachment
		and outward motion of the pearls generated by
		the instability of the ring. The three-dimensional view highlights the
		loss of rotational symmetry and the sequence of outward excursions of
		these localized structures that characterize the $m=2$ evolution.

		Apart from the quantitative shifts in the thresholds and parameter ranges identified in the time-independent bifurcation diagrams, the saturable nonlinearity \eqref{nlin:sat} shows no qualitative differences in its time-dependent supratransmission dynamics. Once the drive exceeds the corresponding critical amplitude, the evolution follows the same sequence
		observed in the cubic--quintic model: an initial evanescent response, formation of a driven standing state near the inner boundary, emission of pearls, and their subsequent propagation and deformation in the two-dimensional domain. 

		\section{Nonlinear dynamics in the allowed band}\label{sec:allowed}
		
		\begin{figure*}[thbp!]
			\centering
			{\includegraphics[scale=.7]{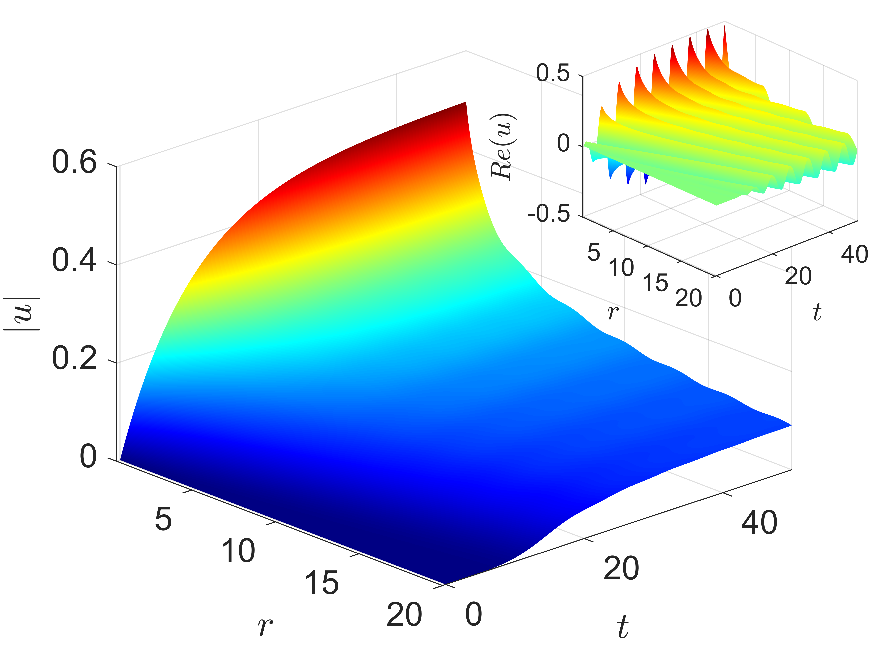}\label{subfig:TD_Omega_-1_m_0_local_combine}}\quad
			\caption{{
					Spatiotemporal evolution of the radially averaged amplitude $|u(r,t)|$ in the cubic--quintic model for $\Omega = -1$. The solution is axisymmetric. The main panel shows the propagation of waves away from the driven inner boundary $r = r_{\min}$, illustrating smooth outward radiation in the allowed band. The inset displays $\mathrm{Re}(u)$, confirming the effectiveness of the transparent boundary condition (TBC) at $r = r_{\max}$. Parameters: $F_0 = 0.5$, $r_{\min} = 0.5$, $s = 0.02$, and $m = 0$.
				}
			}
			\label{fig:TD_Omega_-1_m_0}
		\end{figure*}
		
		\begin{figure*}[thbp!]
			\centering
			\subfloat[$t=0$]{\includegraphics[scale=.45]{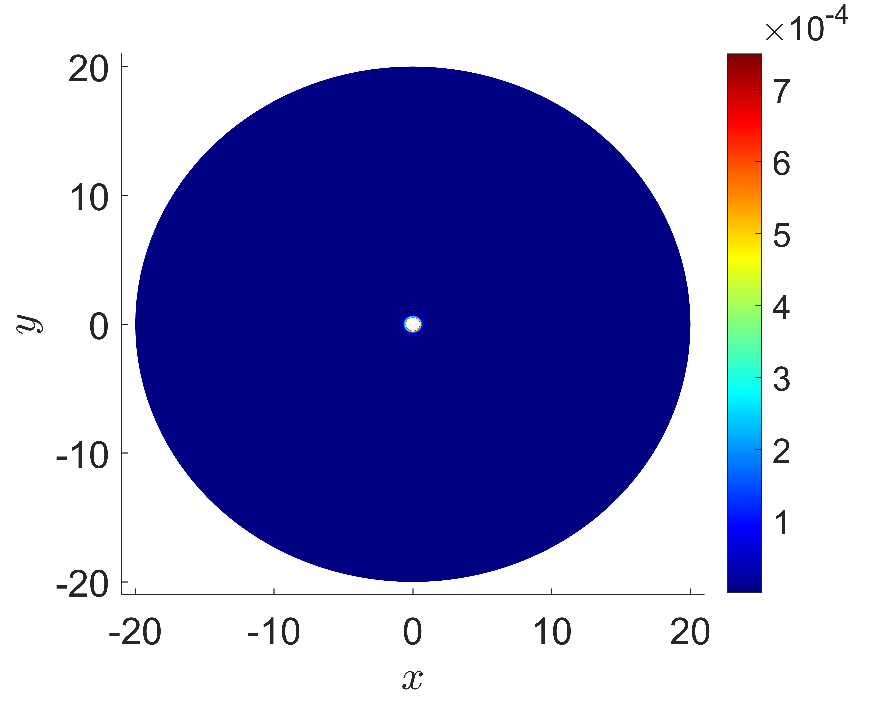}\label{subfig:TD_CQ_F0=0_75_s=0_02_m=0_q=128_rm=0_5_t=0}}
			\subfloat[$t=10$]{\includegraphics[scale=.45]{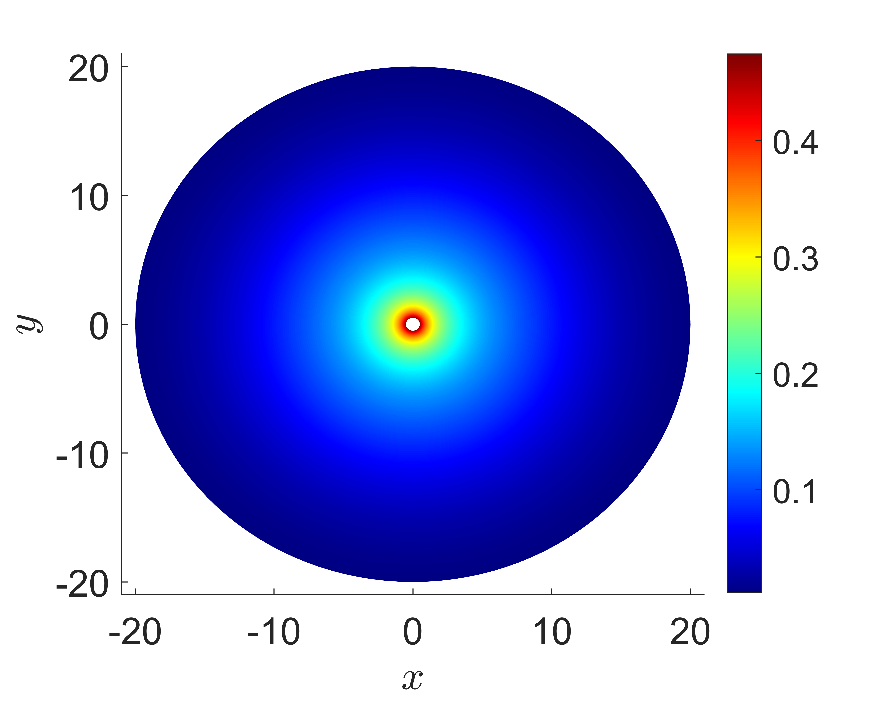}\label{subfig:TD_CQ_F0=0_75_s=0_02_m=0_q=128_rm=0_5_t=10}}\\
			\subfloat[$t=100$]{\includegraphics[scale=.45]{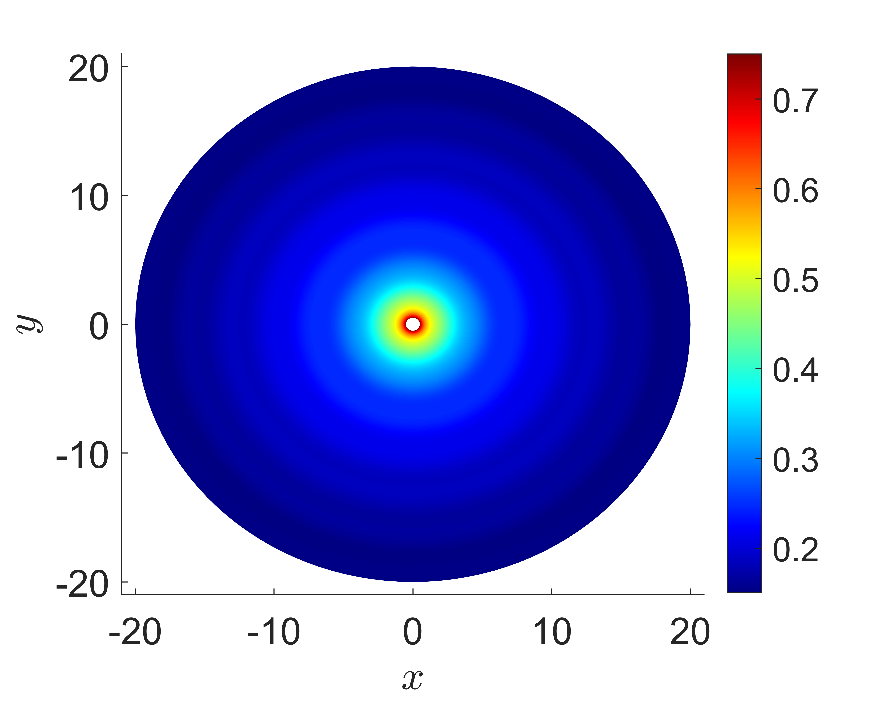}\label{subfig:TD_CQ_F0=0_75_s=0_02_m=0_q=128_rm=0_5_t=100}}
			\subfloat[$t=160$]{\includegraphics[scale=.45]{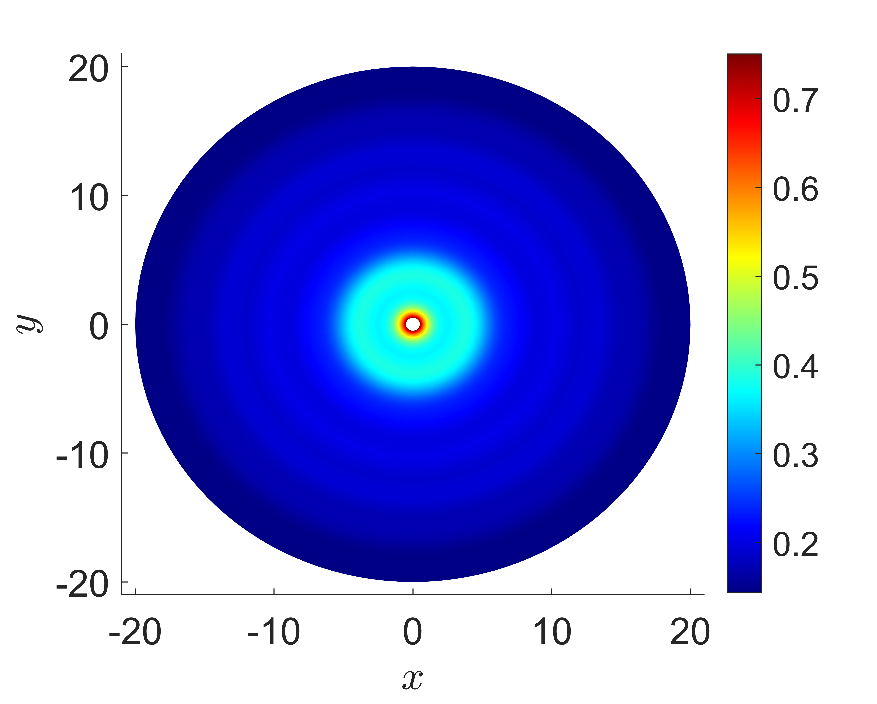}\label{subfig:TD_CQ_F0=0_75_s=0_02_m=0_q=128_rm=0_5_t=160}}\\
			\subfloat[$t=179$]{\includegraphics[scale=.45]{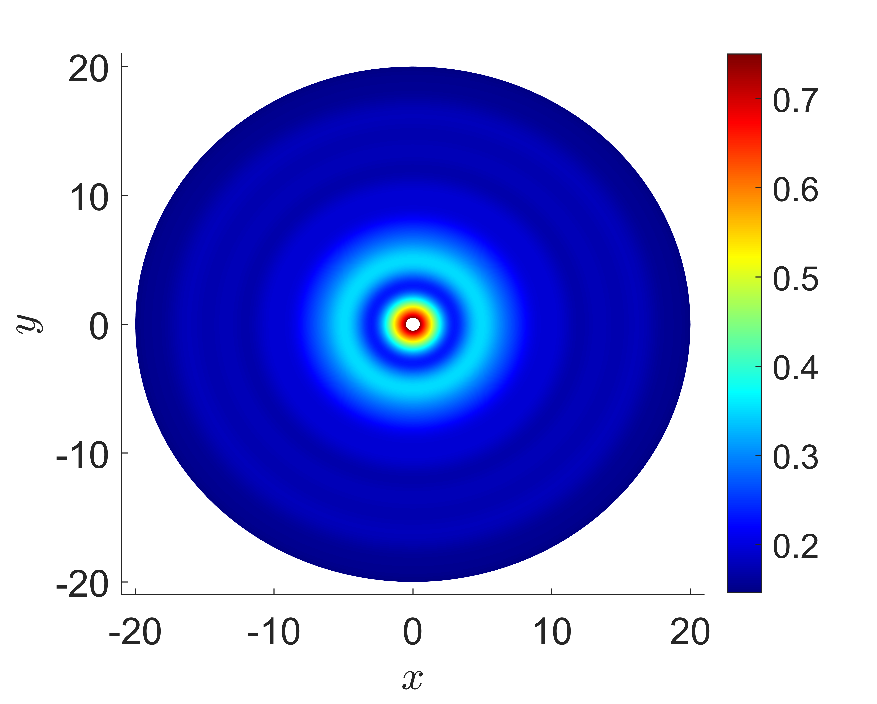}\label{subfig:TD_CQ_F0=0_75_s=0_02_m=0_q=128_rm=0_5_t=179}}
			\subfloat[$t=209$]{\includegraphics[scale=.45]{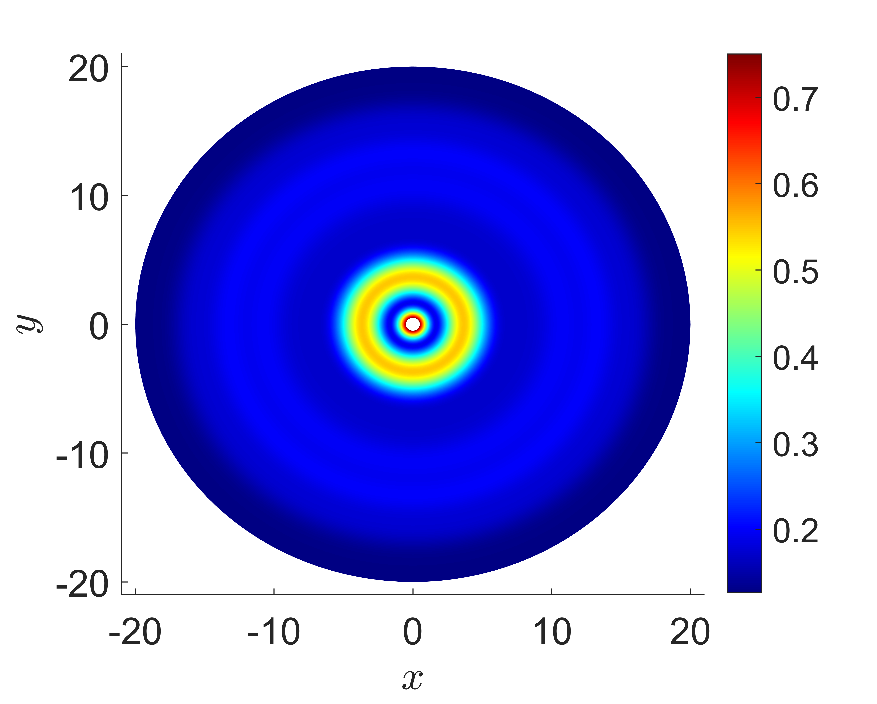}\label{subfig:TD_CQ_F0=0_75_s=0_02_m=0_q=128_rm=0_5_t=209}}
			\caption{
				Time evolution of $|u(x,y,t)|$ for $\Omega = -1$ with $F_0 = 0.75$, showing the emergence of nonlinear radial structures in the allowed band. Panels (a)--(f) correspond to increasing times. The initially smooth radiative field develops concentric ring patterns that undergo repeated expansion and contraction, indicating a transition from linear propagation to nonlinear, structured dynamics on a non-vanishing background. This is akin to a supratransmission with a frequency in the allowed band. Other parameters are as in Fig.~\ref{fig:TD_Omega_-1_m_0}.
			}
			\label{fig:TD_CQ_F0=0_75_s=0_02_m=0_q=128_rm=0_5}
		\end{figure*}

		\begin{figure*}[thbp!]
			\centering
			{\includegraphics[scale=.7]{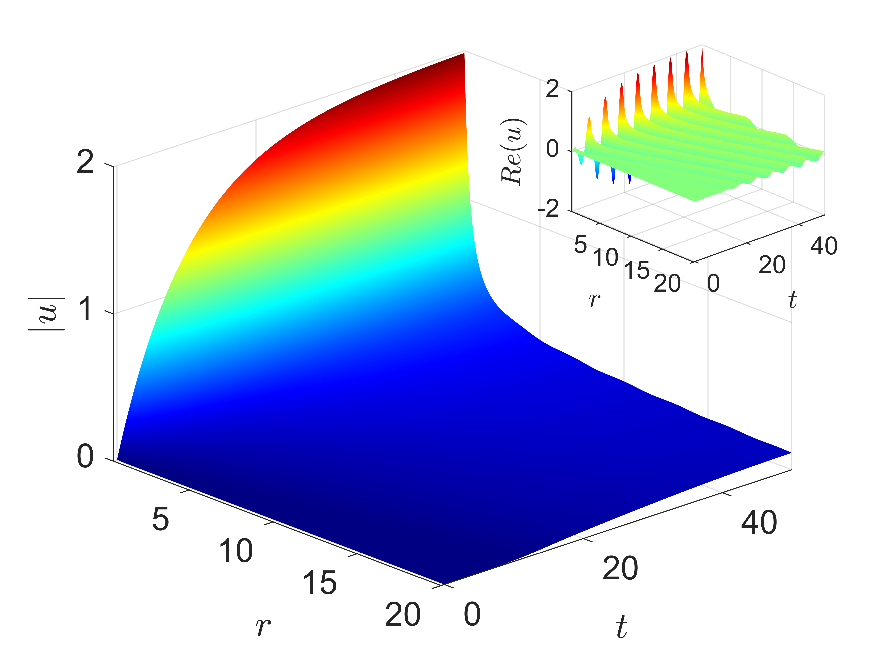}}
			\caption{
				Spatiotemporal evolution in the allowed band for nonzero azimuthal charge $m = 2$ and subcritical amplitude $F_0 = 2$. The solution clearly has no vanishing tail. The other parameters are $r_{\min} = 0.5$ and $s = 0.02$.
			}
			\label{fig:TD_Omega_-1_m_2_local}
		\end{figure*}
		
		\begin{figure*}[thbp!]
			\centering
			\subfloat[$t=0$]{\includegraphics[scale=.45]{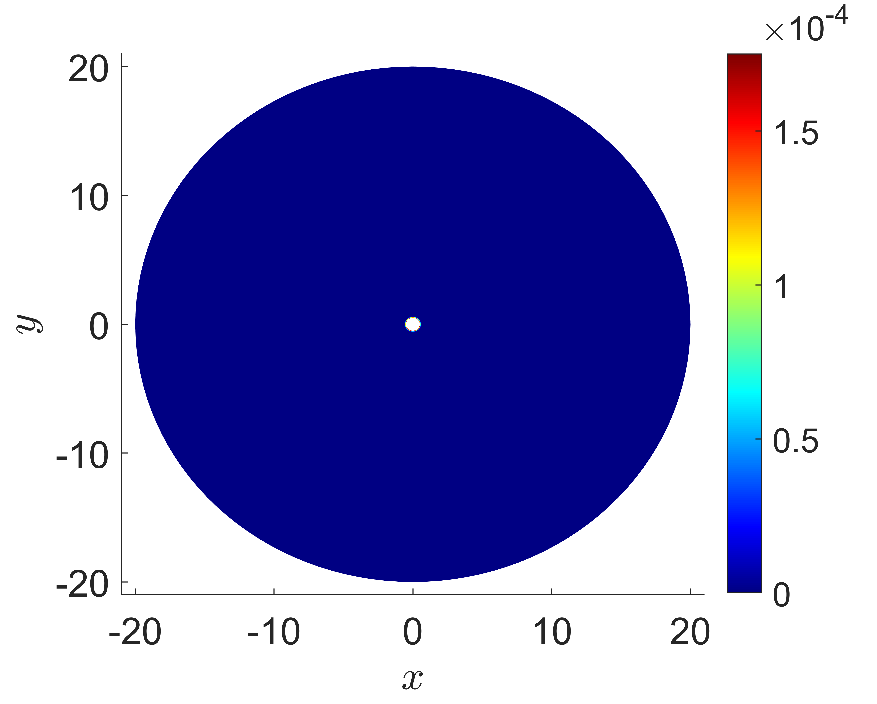}\label{subfig:TD_CQ_F0=3_5_s=0_02_m=2_q=128_rm=0_5_t=0}}
			\subfloat[$t=10$]{\includegraphics[scale=.45]{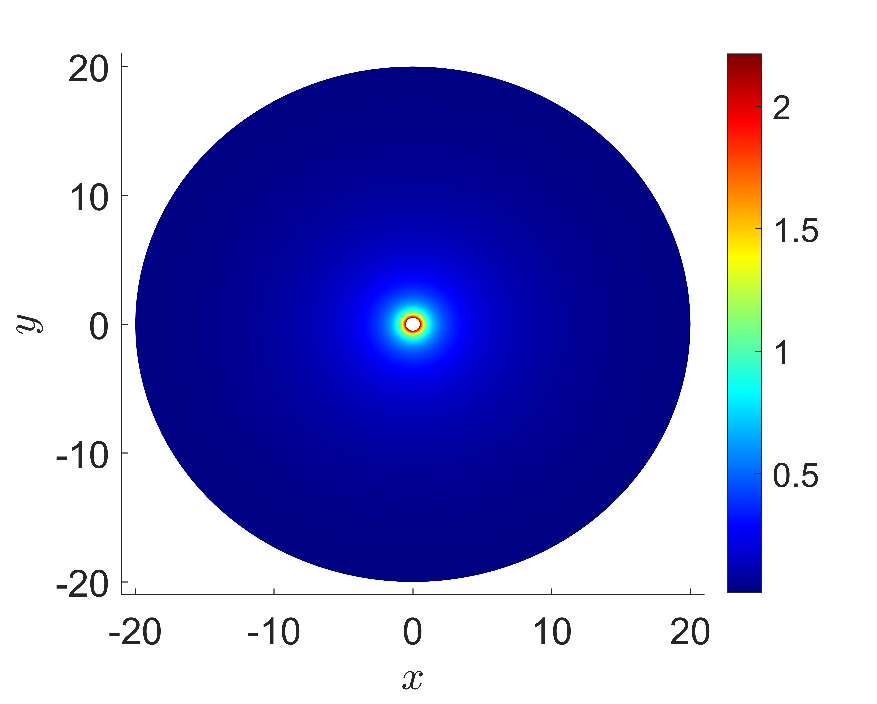}\label{subfig:TD_CQ_F0=3_5_s=0_02_m=2_q=128_rm=0_5_t=10}}\\
			\subfloat[$t=15$]{\includegraphics[scale=.45]{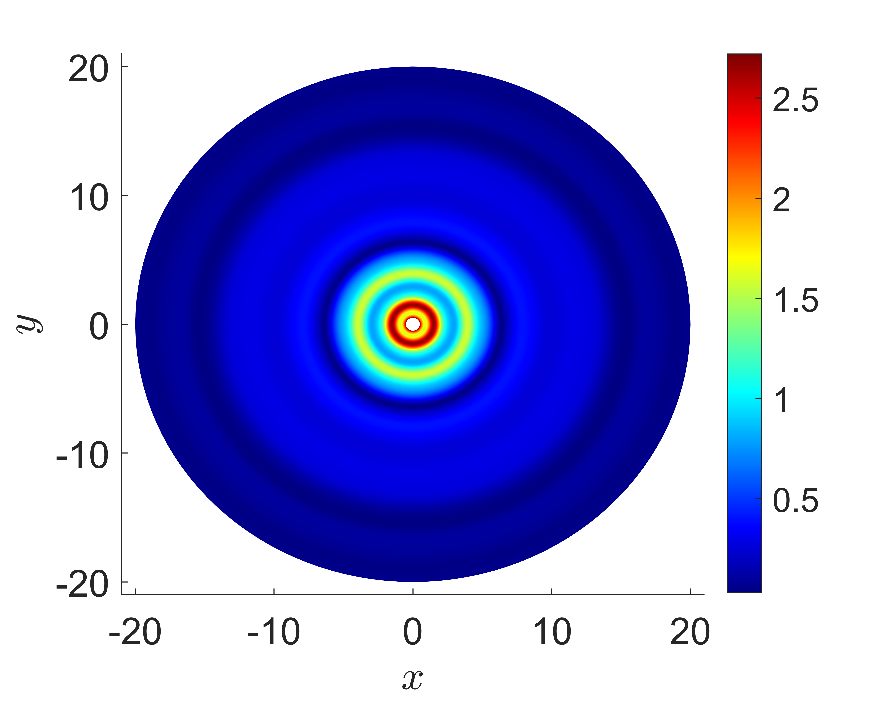}\label{subfig:TD_CQ_F0=3_5_s=0_02_m=2_q=128_rm=0_5_t=15}}
			\subfloat[$t=16$]{\includegraphics[scale=.45]{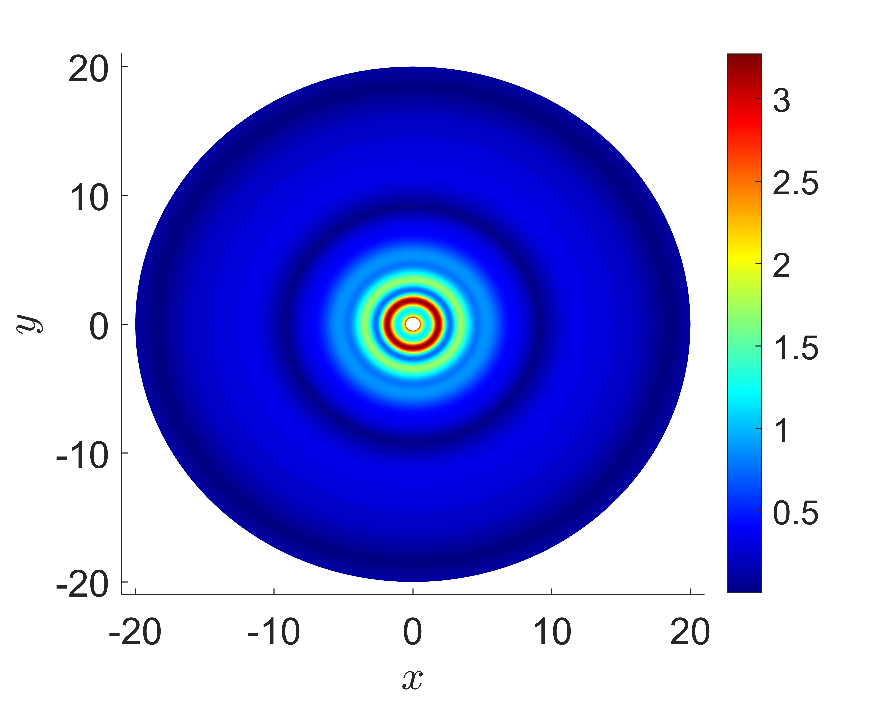}\label{subfig:TD_CQ_F0=3_5_s=0_02_m=2_q=128_rm=0_5_t=16}}\\
			\subfloat[$t=17$]{\includegraphics[scale=.45]{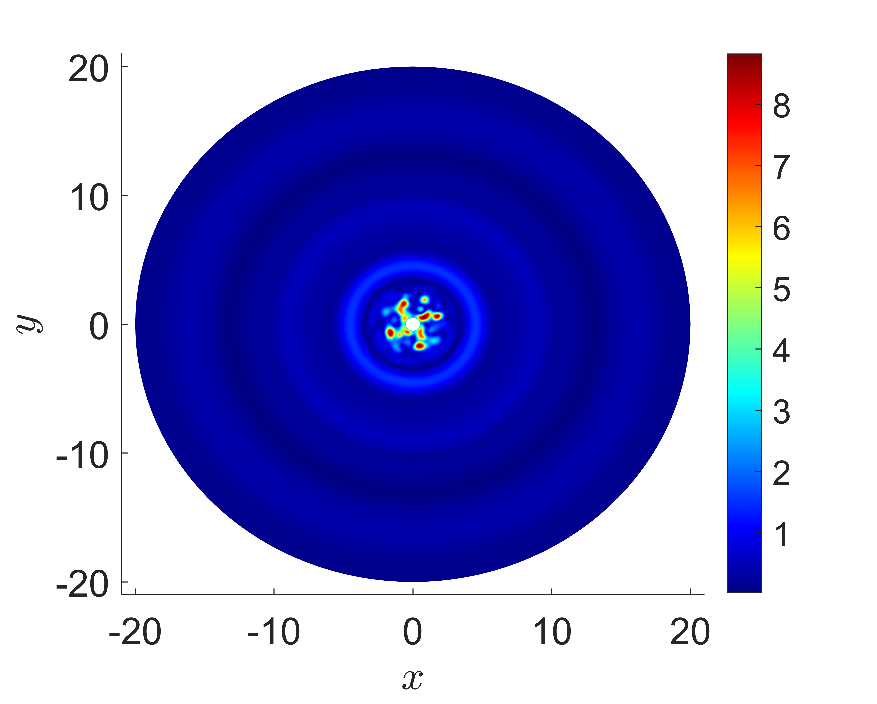}\label{subfig:TD_CQ_F0=3_5_s=0_02_m=2_q=128_rm=0_5_t=17}}
			\subfloat[$t=19$]{\includegraphics[scale=.45]{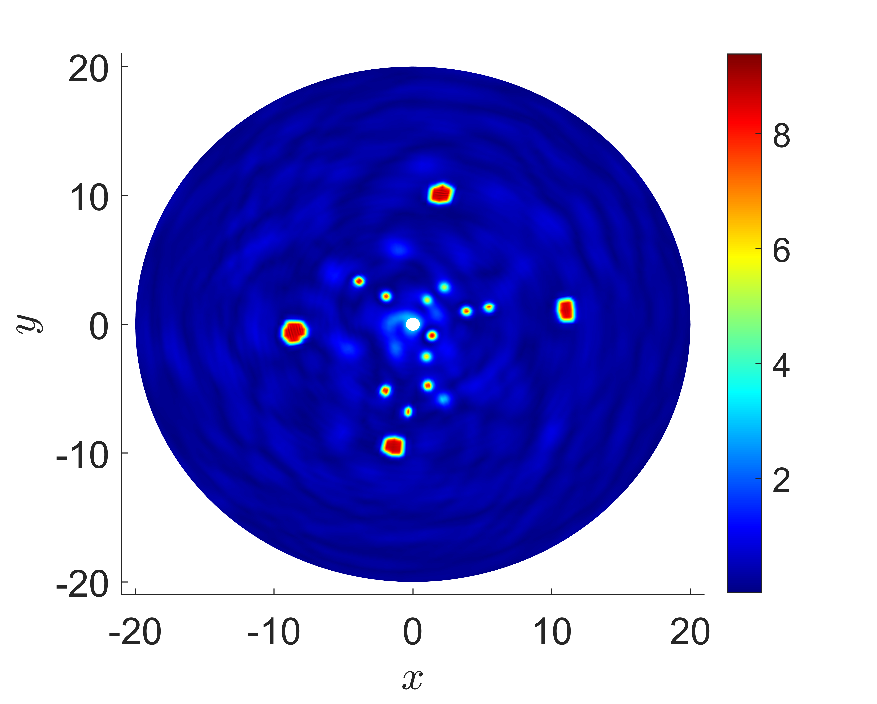}\label{subfig:TD_CQ_F0=3_5_s=0_02_m=2_q=128_rm=0_5_t=19}}
			\caption{
				Time evolution for $m = 2$ with $F_0 = 3.5$, illustrating symmetry breaking in the allowed band. Panels (a)–(f) show the transition from an initially axisymmetric state to a configuration with multiple localized structures (pearls) generated by azimuthal instability. These structures propagate outward on top of a persistent radiative background. {The remaining parameters are as in Fig}.~\ref{fig:TD_Omega_-1_m_2_local}. Compared to Fig.~\ref{fig:TD_CQ_F0=10_s=0_02_m=2_q=0_rm=0_5}, we observe several big pearls.
			}
			\label{fig:TD_CQ_F0=3_5_s=0_02_m=2_q=128_rm=0_5}
		\end{figure*}

		We now examine the time-dependent dynamics of Eqs.~\eqref{main1}--\eqref{drive} when the driving frequency lies in the allowed band, i.e., $\Omega < 0$, where linear waves are propagating. In this regime, boundary forcing generates outward-traveling radiation even for arbitrarily small amplitudes. Hence, unlike the band-gap case studied in Secs.~\ref{Sec:Stand} and \ref{sec:dynamics}, energy transmission is not forbidden at the linear level. Nevertheless, the nonlinear response exhibits a distinct amplitude-dependent transition that resembles supratransmission, despite occurring within the allowed band.
		
		As the solution we consider in this section does not decay to a negligible amplitude as $r\to\infty$, we use an alternative numerical setup. We solve Eqs.~\eqref{main1}--\eqref{drive} using a semi-implicit Crank--Nicolson scheme for the linear part, while treating the nonlinear term explicitly. The resulting method is second-order accurate in time and in space for the linear operator, but the explicit	nonlinearity imposes a stability restriction on the time step. At the outer boundary at $r=r_{\max}$, we impose a transparent boundary condition (TBC) based on the Dirichlet-to-Neumann map and formulated as a fractional, nonlocal operator in time (see Appendix~\ref{sec:TBC}). This condition is exact	for the one-dimensional linear Schr\"odinger equation. It allows outgoing waves to leave the computational domain without spurious reflections and is essential for resolving the long-time dynamics in the propagating regime. The adiabatic ramp in \eqref{eq:inc} is implemented with $\tau=10$.
		
		Figure~\ref{fig:TD_Omega_-1_m_0} shows the spatiotemporal evolution of the radially averaged amplitude $|u(r,t)|$ for a representative case with $\Omega = -1$, $F_0 = 0.5$, $r_{\min} = 0.5$, $s = 0.02$, and $m = 0$. The boundary forcing generates waves that propagate smoothly away from $r = r_{\min}$ into the bulk. The amplitude grows gradually and no localization is observed, indicating a predominantly linear response. The inset displaying $\mathrm{Re}(u)$ confirms that the TBC effectively absorbs outgoing radiation.
		
		A qualitatively different behavior emerges when the driving amplitude is increased above a threshold. Figure~\ref{fig:TD_CQ_F0=0_75_s=0_02_m=0_q=128_rm=0_5} shows the evolution for $F_0 = 0.75$, with all other parameters unchanged. Although linear propagation persists, the system develops pronounced radial structures. Instead of a purely radiative field, concentric ring patterns form and undergo repeated expansion and contraction. These structures resemble the oscillatory ring solitons observed in the band-gap regime (cf.~Figs.~\ref{fig:TD_CQ_F0=10_s=0_02_m=0_q=0_rm=0_5} and \ref{subfig:isosurface_CQ_F0=1_5_s=0_02_m=0_rm=0_5}), but now appear on top of a non-vanishing background. This indicates that above a certain amplitude, nonlinear effects reorganize the energy into coherent radial patterns even though linear propagation is allowed. The transition is associated with the disappearance of nearby steady states, analogous to the saddle-node mechanism discussed in Sec.~\ref{Sec:Stand}, but embedded within a radiative background.
		
		The role of azimuthal perturbations is illustrated in Figs.~\ref{fig:TD_Omega_-1_m_2_local} and \ref{fig:TD_CQ_F0=3_5_s=0_02_m=2_q=128_rm=0_5}. For $m = 2$ and a moderate amplitude $F_0 = 2$ (Fig.~\ref{fig:TD_Omega_-1_m_2_local}), the solution remains largely axisymmetric. The field consists of outward-propagating waves with weak angular modulation, and no strong localization occurs. This corresponds to a subcritical regime in which nonlinear effects are insufficient to destabilize the radial structure.
		
		When the amplitude is increased further to $F_0 = 3.5$, a clear transition takes place (Fig.~\ref{fig:TD_CQ_F0=3_5_s=0_02_m=2_q=128_rm=0_5}). The initially axisymmetric wave becomes unstable to azimuthal perturbations, leading to symmetry breaking and the formation of multiple localized structures (pearls). These structures detach from the inner boundary and propagate outward, similar to the dynamics observed in Fig.~\ref{fig:TD_CQ_F0=10_s=0_02_m=2_q=0_rm=0_5} for the forbidden band. However, here the pearls evolve on top of a persistent radiative background. The coexistence of localized excitations and extended waves shows the absence of a band gap and the hybrid nature of the dynamics in this regime.
		
		Taken together, these results show that although linear transmission is always present for $\Omega < 0$, the nonlinear system exhibits an amplitude threshold that marks a transition from smooth radiative propagation to structured, and eventually symmetry-breaking, dynamics. This transition is akin to the supratransmission threshold in the band gap, even though it is in the strict classical sense, since energy transport is not forbidden below it. 
		
		Bifurcation analysis of standing waves in the allowed band is not straightforward. The main difficulty is that the appropriate boundary condition at the outer radius $r = r_{\max}$ for the time-independent problem is not known (see \cite{derkach2025transparent} for a recent discussion on TBCs in stationary linear equations). Since solutions in the allowed band correspond to propagating waves, they do not decay as $r \to \infty$, and any truncation of the domain necessarily requires an artificial boundary condition. Standard choices introduce reflections, which modify the solution profiles. Consequently, the resulting bifurcation diagrams and the inferred threshold amplitudes depend sensitively on both the imposed boundary condition and the size of the computational domain. This lack of a reliable outer boundary condition prevents a meaningful continuation of standing-wave branches analogous to that carried out in Sec.~\ref{Sec:Stand}.
		
		\section{Conclusions}
		\label{Sec:Conclusions}
		
		We investigated supratransmission in a two-dimensional nonlinear
		Schr\"odinger equation with a central hole driven harmonically along the
		inner boundary. The setting combines nonlinearity, geometry, and
		azimuthal forcing, providing a natural extension of the classical
		one-dimensional supratransmission scenario to higher dimensions.
		
		The bifurcation analysis of standing waves showed that the critical
		driving amplitude depends on the inner radius $r_{\min}$ and the
		azimuthal charge $m$. Small values of $r_{\min}$ enhance radial confinement
		and raise the threshold for standing-wave disappearance, while larger
		$r_{\min}$ produce broader profiles and smoother bifurcation curves. The
		cubic--quintic and saturable models share the same qualitative structure,
		but differ quantitatively: the cubic--quintic case exhibits more
		pronounced turning points, whereas the saturable nonlinearity yields a
		weaker dependence on $r_{\min}$ and $m$.
		
		A variational approximation based on a radial ansatz reproduced the
		dependence of the critical drive on $r_{\min}$ and $m$ with good accuracy for
		moderate azimuthal charge. Its agreement with the numerical bifurcation
		curves clarifies how the nonlinear response determines the onset of
		supratransmission and the shape of the stationary states near the turning
		point.
		
		Time-dependent simulations in the band-gap regime confirmed these
		trends. For $m=0$, the transition beyond the threshold occurs through the
		formation and repeated excursion of localized radial pulses. For nonzero
		azimuthal charge, the driven state loses axisymmetry more rapidly, and
		pearls, i.e., localized two-dimensional structures, detach from the
		boundary and propagate outward. The presence of angular momentum
		therefore accelerates symmetry breaking and produces a richer sequence
		of outgoing excitations.
		
		Isosurface visualizations provided a complementary view of these
		processes by highlighting the outward motion and deformation of the
		emitted structures in $(r,\theta,t)$ space. These plots capture the
		departure from axisymmetry and the organization of the radial and angular
		excursions across time.
		
		{ We also examined the dynamics in the allowed band, where linear waves are
			propagating. In this regime, energy transmission occurs for arbitrarily
			small driving amplitudes; however, the nonlinear system still exhibits a
			clear amplitude-dependent transition. Below this transition, the response
			is dominated by smooth outward radiation. Above it, nonlinear effects
			organize the field into coherent structures, including oscillatory ring
			patterns and, for nonzero azimuthal charge, symmetry-breaking states
			leading to the formation of pearls on top of a persistent radiative
			background. This behavior can be interpreted as an in-band analogue of
			supratransmission, although it does not correspond to a strict threshold
			for transmission in the forbidden band. }
		
		Overall, the results show that supratransmission in this two-dimensional
		geometry retains the qualitative features of the one-dimensional
		mechanism, while exhibiting additional structure due to curvature and
		azimuthal forcing. The extension to the allowed-band regime further shows
		that similar nonlinear transitions persist even when linear propagation
		is present. These findings may inform the design of driven nonlinear
		media where energy injection and localization depend sensitively on
		geometry and topological charge.		
		
		
		Future work could extend these findings to three-dimensional systems, where the additional spatial degree of freedom would enable a richer set of bifurcation phenomena and turbulence dynamics \cite{kartashov2011solitons,mihalache2021localized}. {Extension to genuinely discrete two-dimensional lattice systems, where bandgap structures arise from lattice periodicity, also constitutes an interesting direction for future investigation \cite{lieb1989two,li2022topological}.} Experimental validation of these findings in, e.g., disk-shaped Josepshon junctions \cite{gurlich2010visualizing,ahmad2010existence,castro2020stability}, optical waveguides or Bose-Einstein condensates will be another interesting direction. Finally, integrating machine learning techniques to analyze high-dimensional bifurcation structures could provide innovative tools for exploring complex nonlinear systems \cite{raissi2019physics,shahab2025physics}.
		

		\section*{Acknowledgment}
		R.K. acknowledges that this research is funded by the ITB Research Program 2026 under the ITB International Research Scheme through the Directorate of Research and Innovation, Institut Teknologi Bandung (Project ID: FMIPA.PN-6-126-2026).
		The authors acknowledge the two referees for their support and remarks. 
		
		\appendix
		\section{Transparent Boundary Conditions}
		\label{sec:TBC}
		
		We consider the free two-dimensional Schr\"odinger equation posed on the
		annular domain $r\in(r_{\min},r_{\max})$, $\theta\in[0,2\pi)$:
		\begin{equation}
			i\partial_t u
			=
			\partial_{rr}u+\frac{1}{r}\partial_r u
			+\frac{1}{r^2}\partial_{\theta\theta}u,
			\qquad t>0.
			\label{eq:schrodinger}
		\end{equation}
		Our goal is to construct a transparent boundary condition (TBC) at
		$r=r_{\max}$ that exactly reproduces the outgoing solution in the
		exterior domain $r>r_{\max}$. TBCs for Schr\"odinger-type
		equations have been extensively studied; see, e.g., Refs.~\cite{antoine2008review,arnold2003discrete}.
		
		\subsection{Modal decomposition and Laplace-domain formulation}
		
		Expanding the solution in angular Fourier modes,
		\begin{equation}
			u(r,\theta,t)
			=
			\sum_{m\in\mathbb{Z}}U_m(r,t)e^{im\theta},
			\label{eq:fourier}
		\end{equation}
		yields the decoupled radial equations
		\begin{equation}
			i\partial_t U_m
			=
			\partial_{rr}U_m+\frac{1}{r}\partial_r U_m
			-\frac{m^2}{r^2}U_m.
			\label{eq:modal}
		\end{equation}
		
		We apply the Laplace transform in time,
		\[
		\hat U_m(r,s)=\int_0^\infty e^{-st}U_m(r,t)\,dt,
		\qquad \text{Re}(s)>0,
		\]
		and assume that the initial data vanish in the exterior domain. Then
		\eqref{eq:modal} becomes
		\begin{equation}
			\hat U_m''+\frac{1}{r}\hat U_m'
			-
			\left(\frac{m^2}{r^2}+is\right)\hat U_m=0.
			\label{eq:laplace}
		\end{equation}
		
		Introducing $\kappa=(is)^{1/2}$ (principal branch, $\text{Re}(\kappa)>0$) and
		$\xi=\kappa r$, Eq.~\eqref{eq:laplace} reduces to the modified Bessel equation
		\begin{equation}
			\xi^2 \frac{d^2 \hat U_m}{d\xi^2}
			+
			\xi \frac{d \hat U_m}{d\xi}
			-
			(\xi^2+m^2)\hat U_m=0.
			\label{eq:bessel}
		\end{equation}
		Its general solution is a linear combination of $I_m(\kappa r)$ and
		$K_m(\kappa r)$. Since $I_m$ grows exponentially as $\text{Re}(\kappa)r\to\infty$,
		the radiation condition selects the decaying solution
		\begin{equation}
			\hat U_m(r,s)=B_m(s)\,K_m(\kappa r),
			\qquad r\ge r_{\max}.
			\label{eq:exterior}
		\end{equation}
		
		Differentiating \eqref{eq:exterior} and eliminating $B_m(s)$ yields the
		Dirichlet-to-Neumann (DtN) map
		\begin{equation}
			\partial_r \hat U_m\big|_{r_{\max}}
			=
			\kappa\,\frac{K_m'(\kappa r_{\max})}{K_m(\kappa r_{\max})}
			\,\hat U_m\big|_{r_{\max}}.
			\label{eq:exactTBC}
		\end{equation}
		Such DtN operators form the basis of exact nonreflecting boundary conditions \cite{antoine2008review,arnold2003discrete}.
		
		\subsection{Large-radius asymptotics}
		
		Let $z=\kappa r_{\max}$. For $|\arg z|<3\pi/2$, the modified Bessel
		function admits the asymptotic expansion~\cite{abramowitz1964handbook}
		\begin{subequations}
			\begin{align}
				K_m(z)
				&\sim
				\sqrt{\frac{\pi}{2z}}\,e^{-z}
				\sum_{k=0}^\infty \frac{a_k(m)}{z^k}, \\
				a_k(m)
				&=
				\frac{\prod_{j=1}^{k}\bigl[4m^2-(2j-1)^2\bigr]}{k!\,8^k}.
			\end{align}
			\label{eq:besselasym}
		\end{subequations}
		
		From this expansion, one obtains
		\begin{equation}
			\frac{K_m'(z)}{K_m(z)}
			=
			-1-\frac{1}{2z}-\frac{4m^2-1}{8z^2}
			+O(z^{-3}),
			\label{eq:ratio_z}
		\end{equation}
		and hence
		\begin{equation}
			\kappa\frac{K_m'(\kappa r_{\max})}{K_m(\kappa r_{\max})}
			=
			-\kappa
			-\frac{1}{2r_{\max}}
			-\frac{4m^2-1}{8\kappa r_{\max}^2}
			+O\!\left(\frac{1}{r_{\max}^3}\right),
			\label{eq:besselratio}
		\end{equation}
		where the remainder is understood at the level of the Laplace-domain symbol. Such asymptotic expansions of DtN operators are classical in wave propagation
		problems \cite{engquist1977absorbing,antoine1999bayliss,antoine2001construction}.
		
		Introduce the angular operator
		\begin{equation}
			\mathcal{A}:=-4\partial_{\theta\theta}-1,
			\qquad
			\mathcal{A}e^{im\theta}=(4m^2-1)e^{im\theta},
			\label{eq:Aop}
		\end{equation}
		and use $\kappa=(is)^{1/2}=e^{i\pi/4}s^{1/2}$. After summing over the angular modes (see (\ref{eq:fourier})), the asymptotic DtN map \eqref{eq:exactTBC} becomes, in Laplace space,
		\begin{equation}
			\partial_r \hat u\big|_{r_{\max}}
			=
			-e^{i\pi/4}s^{1/2}\hat u
			-\frac{1}{2r_{\max}}\hat u
			-\frac{e^{-i\pi/4}}{8r_{\max}^2}
			\mathcal{A}\,s^{-1/2}\hat u
			+\cdots.
			\label{eq:laplace_symbol}
		\end{equation}
		
		Taking the inverse Laplace transform and using the convolution properties yields the time-domain boundary operator
		\begin{equation}
			\begin{aligned}
				\partial_r u\big|_{r_{\max}}
				={}&
				-e^{i\pi/4}D_{RL}^{1/2}u
				-\frac{1}{2r_{\max}}u \\
				&-\frac{e^{-i\pi/4}}{8r_{\max}^2}
				\mathcal{A}I^{1/2}u
				+\cdots,
			\end{aligned}
			\label{eq:TBC}
		\end{equation}
		where $D_{RL}^{1/2}$ denotes the Riemann--Liouville fractional derivative
		and $I^{1/2}$ the fractional integral of order $1/2$. Such time-domain convolution operators are characteristic of Schr\"odinger TBCs \cite{antoine2008review}.
		
		For sufficiently smooth functions,
		\begin{equation}
			I^\alpha u(t)
			=
			\frac{1}{\Gamma(\alpha)}
			\int_0^t (t-s)^{\alpha-1}u(s)\,ds,
			\label{eq:RL_def}
		\end{equation}
		and
		\begin{equation}
			D_{RL}^\alpha u(t)
			=
			\frac{d}{dt}I^{1-\alpha}u(t),
			\qquad 0<\alpha<1,
			\label{eq:RL_deriv}
		\end{equation}
		see, e.g., \cite{podlubny1999fractional}.
		
		If the boundary trace satisfies the compatibility condition
		$u(r_{\max},\theta,0)=0$, then the Riemann--Liouville and Caputo
		derivatives coincide, and the leading term in \eqref{eq:TBC} can be
		written in Caputo form.
		\\
		\subsection{Time discretization of the fractional operators}
		
		Let $t_n=n\Delta t$, $n=0,1,\dots$, be a uniform temporal grid, and
		$u^n\approx u(t_n)$.
		
		\subsubsection{L1 approximation of the half-derivative}
		
		For the Caputo derivative (assuming sufficient regularity),
		\begin{equation}
			D_C^\alpha u(t)
			=
			\frac{1}{\Gamma(1-\alpha)}
			\int_0^t \frac{u'(s)}{(t-s)^\alpha}\,ds,
			\qquad 0<\alpha<1,
			\label{eq:caputo_def}
		\end{equation}
		we split the integral over subintervals $[t_k,t_{k+1}]$ 
		\begin{equation} D_t^\alpha u(t_{n+1}) = \frac{1}{\Gamma(1-\alpha)} \sum_{k=0}^{n} \int_{t_k}^{t_{k+1}}\frac{\partial_s u(s)}{ (t_{n+1}-s)^{\alpha}}\,\,ds, \label{eq:caputo_split_app} \end{equation}
		and approximate $u'(s)$ by a backward difference on each interval 
		\begin{equation} \partial_s u(s) \approx \frac{u^{k+1}-u^k}{\Delta t}, \qquad s\in[t_k,t_{k+1}]. \label{eq:piecewise_diff} \end{equation}
		This yields the classical L1 scheme \cite{sun2006fully,lin2007finite}
		\begin{equation}
			D_C^\alpha u(t_{n+1})
			\approx
			\frac{1}{\Gamma(2-\alpha)(\Delta t)^\alpha}
			\sum_{k=0}^{n}
			b_k\bigl(u^{n+1-k}-u^{n-k}\bigr),
			\label{eq:L1_general_app}
		\end{equation}
		with weights
		\begin{equation}
			b_k=(k+1)^{1-\alpha}-k^{1-\alpha},
			\qquad k\ge0.
			\label{eq:bk_def}
		\end{equation}
		For $\alpha=\tfrac12$, one has $b_k\sim \tfrac12 k^{-1/2}$ as $k\to\infty$,
		reflecting the long-memory nature of the operator. The scheme has
		accuracy $O(\Delta t^{2-\alpha})$ for sufficiently smooth solutions.
		
		\subsubsection{Discrete approximation of the fractional integral}
		
		Evaluating \eqref{eq:RL_def} at $t_{n+1}$ and approximating $u(s)$ by its
		left-endpoint value on each subinterval yields
		\begin{equation}
			I^\alpha u(t_{n+1})
			\approx
			\frac{(\Delta t)^\alpha}{\Gamma(\alpha+1)}
			\sum_{k=0}^{n}
			u^k\Bigl[(n+1-k)^\alpha-(n-k)^\alpha\Bigr].
			\label{eq:L1_integral_app}
		\end{equation}
		
		\medskip
		
		Equations \eqref{eq:L1_general_app} and \eqref{eq:L1_integral_app}
		provide a causal, history-dependent discretization of the fractional
		operators appearing in the asymptotic TBC \eqref{eq:TBC}. The resulting
		boundary condition at time $t_{n+1}$ depends only on $u^{n+1}$ and the
		previous values $\{u^k\}_{k=0}^n$.
		
		An improved transparent boundary condition for the nonlinear Schr\"odinger equation was proposed in Ref.~\cite{zisowsky2008discrete}, based on approximating the nonlinear term as an effective potential.
		
%

	\end{document}